\documentclass{nws}
\expandafter\let\csname equation*\endcsname\relax 
\expandafter\let\csname endequation*\endcsname\relax 
\newtheorem{definition}{Definition}

\usepackage{natbib}
\usepackage{subfigure}
\usepackage{graphicx}
\usepackage{amsmath}
\usepackage{multirow}
\usepackage{epstopdf}
\usepackage{xcolor}
\usepackage{soul}
\usepackage{array}

\newcommand{\hlt}[1]{#1}
\newcommand{\hlc}[1]{#1}

\title[Clustering attributed graphs]
      {Clustering attributed graphs:\\models, measures and methods\footnote{This version has been submitted to the Network Science journal and has been subsequently accepted for publication
      subject to minor revisions. It will appear in a revised form subsequent to peer review and/or editorial input by Cambridge University Press and/or the journal's proprietor (http://journals.cambridge.org/NWS).}}

\author[C.~Bothorel, J.~D.~Cruz, M.~Magnani and B.~Micenkov\'a]
        {CECILE BOTHOREL, JUAN DAVID CRUZ \\
         Department of Logics in Uses, Social Science and Information Science,\\UMR CNRS 3192 Lab-STICC,\\T\'el\'ecom Bretagne, Institut Mines-T\'el\'ecom, Brest, France\\
         }
\email{cecile.bothorel, juan.cruzgomez@telecom-bretagne.eu}  

\author[C.~Bothorel, J.~D.~Cruz, M.~Magnani and B.~Micenkov\'a]{MATTEO MAGNANI\footnote{The author has been partly supported by the Italian Ministry of Education, Universities and Research FIRB grant RBFR107725.}\\
         Computing Science Division, IT Department, Uppsala University, Sweden\\
        }
\email{matteo.magnani@uu.se}  

\author[C.~Bothorel, J.~D.~Cruz, M.~Magnani and B.~Micenkov\'a]{BARBORA MICENKOV\'A\\
         Data Intensive Systems, Department of Computer Science,\\Aarhus University, Denmark
        }
\email{barbora@cs.au.dk}  

\jdate{December 2014}
\pubyear{}
\pagerange{\pageref{firstpage}--\pageref{lastpage}}
\doi{}

\begin{document}

\label{firstpage}

\maketitle

\begin{abstract}
Clustering a graph, i.e., assigning its nodes to groups, is an important operation whose best known application is the discovery of communities in social networks. Graph clustering and community detection have \hlt{traditionally focused on} graphs \hlt{without attributes, with the notable exception of edge weights}. However, \hlt{these models} only provide a partial representation of real social systems, that are thus often \hlt{described} using node attributes, representing features of the \hlt{actors}, and edge attributes, representing different kinds of relationships among them. \hlt{We refer to} these models \hlt{as} \emph{attributed graphs}. Consequently, existing graph clustering methods have been recently extended to deal with node and edge attributes. This article is a literature survey on this topic, organizing and presenting recent research results in a uniform way, characterizing the main existing clustering methods and highlighting their conceptual differences. We also cover the important topic of clustering evaluation and identify current open problems.
\end{abstract}

\tableofcontents

\section{Introduction}\label{sec:intro}


Graphs represent one of the main models to study human relationships. For example, structural properties of social systems can be measured by representing individuals and their relationships as graphs and computing the centrality or prestige of their nodes \citep{Wasserman1994}. Similarly, once a social graph is available, groups of strongly connected individuals (communities) can be identified using clustering algorithms. The application of graphs to the study of social systems motivated and is now a part of a broader discipline called \hlt{\textit{network science}}, focused on the modeling and analysis of relationships between generic entities.
This discipline provides a set of tools (methodologies, methods and measures) to improve our understanding of complex systems, including social and technological environments, transport and communication networks and biological systems. The wide applicability of \hlt{network science} largely relies on the adoption of graph\hlt{-based} models, that thanks to their generality can be applied to a diverse range of scenarios.

\begin{figure}[ht]
\centering
\subfigure[]{
    \includegraphics[width=.28\textwidth]{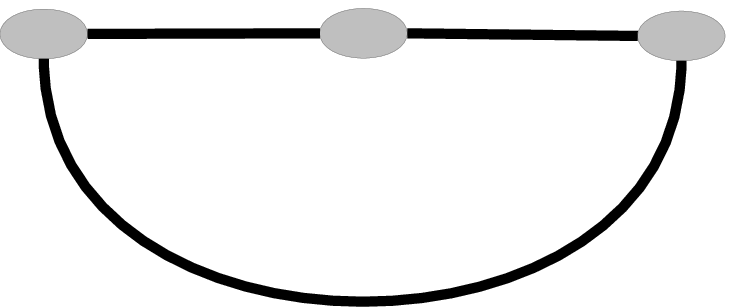}
    \label{fig:simple graph}
}
\subfigure[]{
    \includegraphics[width=.45\textwidth]{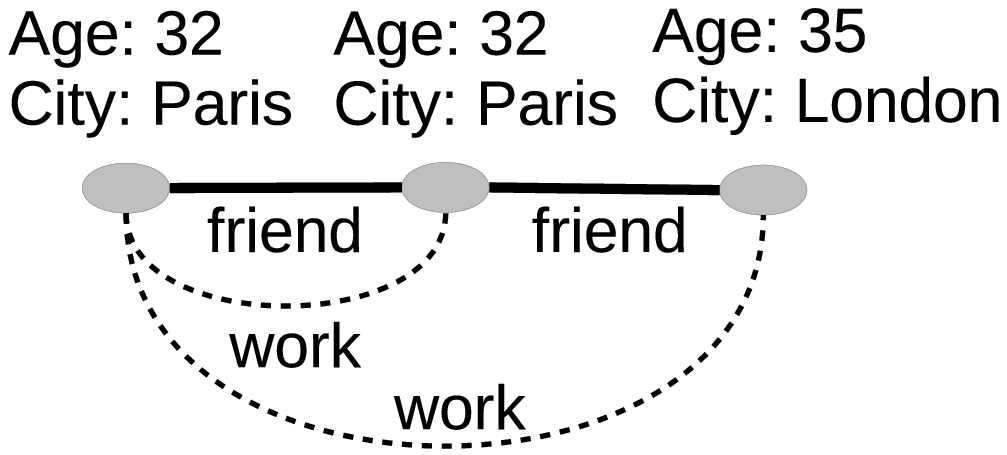}
    \label{fig:attributed graph}
}
\caption{A graph \subref{fig:simple graph} \hlt{provides a simplified} representation of a social system \hlt{which can be easy to understand} but may prevent a deep understanding of its structural and compositional dimensions \subref{fig:attributed graph}}
\label{fig:from simple to attributed graphs}
\end{figure}

However, researchers in social network analysis (SNA) and social sciences have long been aware of the potential value in representing additional information on top of the social graph, and of the potential loss in accuracy when simple nodes and edges are used to represent complex social interactions. For example, according to \citet{Wasserman1994} social networks contain at least three different dimensions: a \textit{structural} dimension corresponding to the social graph, e.g. actors and their relationships, a \textit{compositional} dimension describing the actors, e.g. their personal information, and an \textit{affiliation} dimension indicating group memberships. The existence of multiple relationship types, e.g., \emph{working together}, \emph{being friends} or \emph{exchanging text messages}, has also been studied for a long time, as recently reported by \citet{Borgatti2009}. This last aspect has been referred to as \emph{multiplexity} in the SNA tradition, and can be related to Goffman's concept of \emph{context}, well exemplified by the metaphore of individuals acting on multiple stages depending on their audience \citep{Goffman1974}. As an example, Figure~\ref{fig:attributed graph} highlights how an \emph{attributed} graph may lead to a deeper understanding of social interactions if compared to \hlt{the} corresponding graph \hlt{without attributes} in Figure~\ref{fig:simple graph}.

\subsection{Current trends in attributed graph analysis and mining}

\hlt{A}ttributed graphs have been used for decades to study social environments \hlt{and it has been long recognized that the structure of a social network may not be sufficient to identify its communities }\hlc{\citep{Freeman1996,Hric2014}.} \hlt{However}, recent years have witnessed a renewed attention towards these models, partially motivated by the availability of real data from on-line sources. 
One interesting aspect of real attributed graphs is the observed \emph{dependency} between who the actors are and how they interact, i.e. between the structural and compositional dimensions. For example, \citet{FondNeville2010} have observed the coexistence of social influence and homophily. Social influence states that people who are linked are likely to have similar attributes, thus node attribute values can be interpreted as a result of interactions with other nodes. At the same time, homophily implies that people with similar attributes are likely to build relationships.
These two related phenomena have been observed in real networks by \citet{kossinets2006empirical}, and the dependency between attributes and connectivity has been studied mathematically \citep{KimLeskovec2011}. 

With this in mind, researchers have focused on \emph{attributed graph generators}. Artificially grown graphs are useful to experiment algorithms and run simulations when real data are difficult to collect. They are relevant in testing \emph{what if} scenarios, providing forecasts on future evolutions, and can be used to design graph sampling algorithms when the size of original graphs would otherwise make the analysis impractical \citep{leskovec2005graphs}. 

Prior models, as the well-known preferential attachment mechanism by \citet{Barabasi15101999}, \hlt{have} focused on the social structure. Now the challenge is to generate datasets as close as possible to real-world social graphs, as done by \citet{zheleva:kdd09} where affiliation information is also generated. This model captures previously studied 
properties (e.g. power-law distribution for social degree) but also provides new interesting insights regarding the processes behind group formation.
More recently \citet{DBLP:journals/corr/abs-1112-3265} have proposed a generative \emph{social-attribute network} model based on their empirical observations of Google+ growth. Here attributes
describe user characteristics like name of attended school and group membership. \citet{Du2010,MagnaniSBP2013b} have instead focused on the generation of graphs with interdependent attributes on the edges.

The idea that attributes and connections are generated in an interdependent way has led to the development of specialized analysis methods. 
Several \hlt{graph mining tasks} have been extended to attributed graphs, like link prediction  \citep{Getoor2005Survey,Rossetti2011,DBLP:journals/corr/abs-1112-3265,Sun2012} or attribute inference \citep{Li2009,DBLP:journals/corr/abs-1112-3265,yang2011like}. This survey is dedicated to one of the most relevant and studied operations on graphs and complex networks: graph clustering, often referred to as \emph{community detection} when social graphs are involved. We believe that this is an important and timely effort to facilitate research in this still young area, in particular considering that the discussed approaches have been introduced in different disciplines, often unaware of each other.

\subsection{Clustering attributed graphs}


Although several surveys on graph clustering have been written \citep{Schaeffer2007,Fortunato2010,Aggarwal2010,Coscia2011}, most of the approaches to cluster attributed graphs are more recent and have not been included in these works. At the same time, there is a large literature on (multi-dimensional) clustering of tabular data \citep{Moise2009,Han2011}, but existing surveys in this area have not addressed extensions for graph data. 
Attributed graph clustering can be seen as the confluence of these two fields, the former focusing on the structural and the latter on the compositional aspects.
In this article we focus on recent works resulting from this promising combination.

\begin{table}
\caption{\hlt{Terminology used in this article and synonyms used in the literature}}
\label{tab:terminology}
{\setlength{\tabcolsep}{.1cm}
\begin{tabular}{l m{4cm} m{5cm}}
\hline 
\textbf{main term} & \textbf{synonyms} & \textbf{meaning} \\ 
\hline 
Node & Vertex\hlt{, site, actor} & Basic component of a graph. As an example, a node may indicate that a user has an account on the social media site whose social network is represented by that graph.\\%
\hline 
Edge & Link, arc, tie, connection\hlt{, bond, relation(ship)} & A relationship between two nodes, e.g., a \emph{following} relationship between two Twitter accounts. \hlt{When there is an edge between two nodes we say that they are directly connected.}\\ 
\hline
\hlt{G}raph & Network, social network, layer & A graph without attributes, neither on nodes nor on edges, with the exception of an optional numerical weight on edges indicating the strength of the connection. Edges may be directed or indirected. \\
\hline 
Edge-attributed graph & Multiplex network, multi-layer graph, multidimensional network, edge-labeled multi-graph & Attributes indicate connections of different kinds or inside different graphs. With this term we do not indicate the presence of weights, in which case we explicitly talk of weigh\hlt{t}ed graph/edges. \\ 
\hline 
Node-attributed graph & Node-labeled graph, graph with feature vectors & A feature vector is associated with each node and contains information about it, e.g., age, nationality, language, income. \\ 
\hline 
Attributed graph & Attribute graph, social and affiliation network, relational data, multidimensional network & An edge-attributed graph, or a node-attributed graph, or both. \\ 
\hline 
\hlt{Layer} & \hlt{Aspect, dimension} & \hlt{Sometimes all the edges with the same attribute value in an edge-attributed graph are indicated as a \emph{layer}, e.g., the \emph{Facebook friendship}, \emph{spacial proximity}, \emph{Twitter following}, \emph{colleague} or \emph{family} layers in an attributed graph indicating different types of social relationships.} \\
\hline 
Clustering & Community structure & Assignment of each node to one or more groups of nodes, called clusters. Different criteria can be used to determine whether two nodes should belong to the same cluster.\\
\hline
Partition & Non-overlapping clustering & A clustering where each node is assigned to exactly one cluster.\\
\hline 
\end{tabular}}
\end{table}

The article is organized in three main parts: a review of methods for edge-attributed graphs, a review of methods for node-attributed graphs, and a section on \hlt{practical issues including} the evaluation of clusterings \hlt{and the applicability of different approaches}. We conclude by summarizing the status of the research and discussing the open problems that are more promising according to our view of the area.
Attributed graph clustering has been independently studied in different disciplines, therefore it is important to \hlt{know how different terms have been used in the literature}. In Table~\ref{tab:terminology} we have indicated and briefly explained the terms used in this article.

\section{Clustering edge-attributed graphs}\label{sec:edges}

One way to extend a graph model and to provide additional information to the clustering algorithm is to represent the different kinds of edges among individuals. As an example, in Figure~\ref{fig:from simple to attributed graphs}(b) we can see that the relationship between the two left-most nodes consists of a \emph{friendship} and a \emph{working} edge.


\begin{figure}[ht]
\centering
\subfigure[]{
    \includegraphics[width=.25\textwidth]{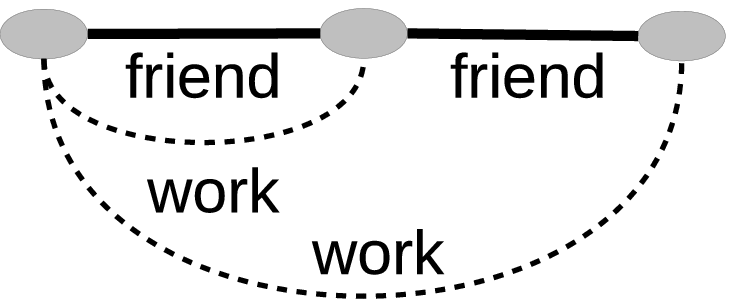}
    \label{fig:dm_multilayer}
}
\subfigure[]{
    \includegraphics[width=.32\textwidth]{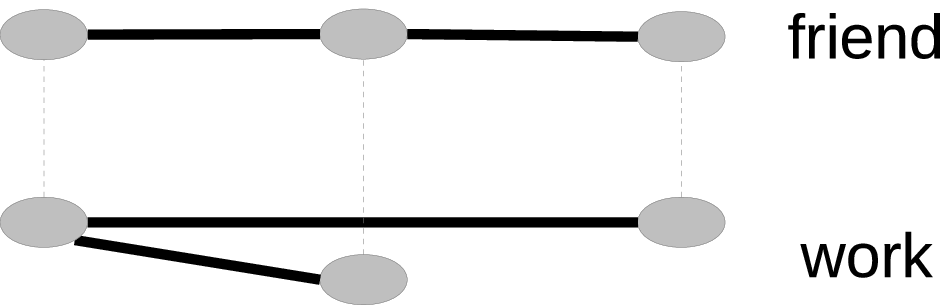}
    \label{fig:dm_flattening}
}
\caption{Two alternative representations of the different edge types in a multigraph}
\label{fig:alternative rep}
\end{figure}

Different models have been used to represent this scenario \citep{minor1983new,lazega1999multiplexity,skvoretz2007reciprocity,Kazienko2010,Berlingerio2011}, sometimes emphasizing the different roles played by individuals with respect to different networks \citep{MagnaniASONAM2011}\hlt{,} including different kinds of nodes \citep{Cai2005b} \hlt{or providing a more general data model to mathematically represent a graph with attributes on both nodes and edges }\hlc{\citep{Kivela2013}}. In Figure~\ref{fig:alternative rep} we can see two alternative representations of the same data, as a multigraph (a) and as a set of interconnected graphs (b). The former\hlt{, sometimes referred to as a \emph{multiplex network},} focuses on a single set of \hlt{nodes} that may have complex relationships between them: 

\begin{definition}[Multi-relational edge-attributed graph]
Given a set of \hlt{nodes $N$} and a set of labels $L$, an edge-attributed graph is a triple $\{G = (V,E,l)\}$ where $V \subseteq$ \hlt{$N$, $(V,E)$ is a multi-graph} and $l: E \rightarrow L$. Each edge $e \in E$ in the graph has an associated label $l(e)$.
\end{definition}

The latter emphasizes how the same \hlt{node} can belong to \hlt{multiple} (social) graphs, \hlt{also known as \emph{layers}}:

\begin{definition}[Multi-layer edge-attributed graph]
Given a set of \hlt{nodes $N$} and a set of labels $L$, an edge-attributed graph is defined as a set of graphs $G_i = (V_i,E_i)$ where $V_i \subseteq \hlc{N}$, $E_i \subseteq V_i \times V_i$. Each graph $G_i$ has an associated unique name $l_i \in L$.
\end{definition}

\begin{figure}[ht]
\centering
\subfigure[]{
    \includegraphics[width=.32\textwidth]{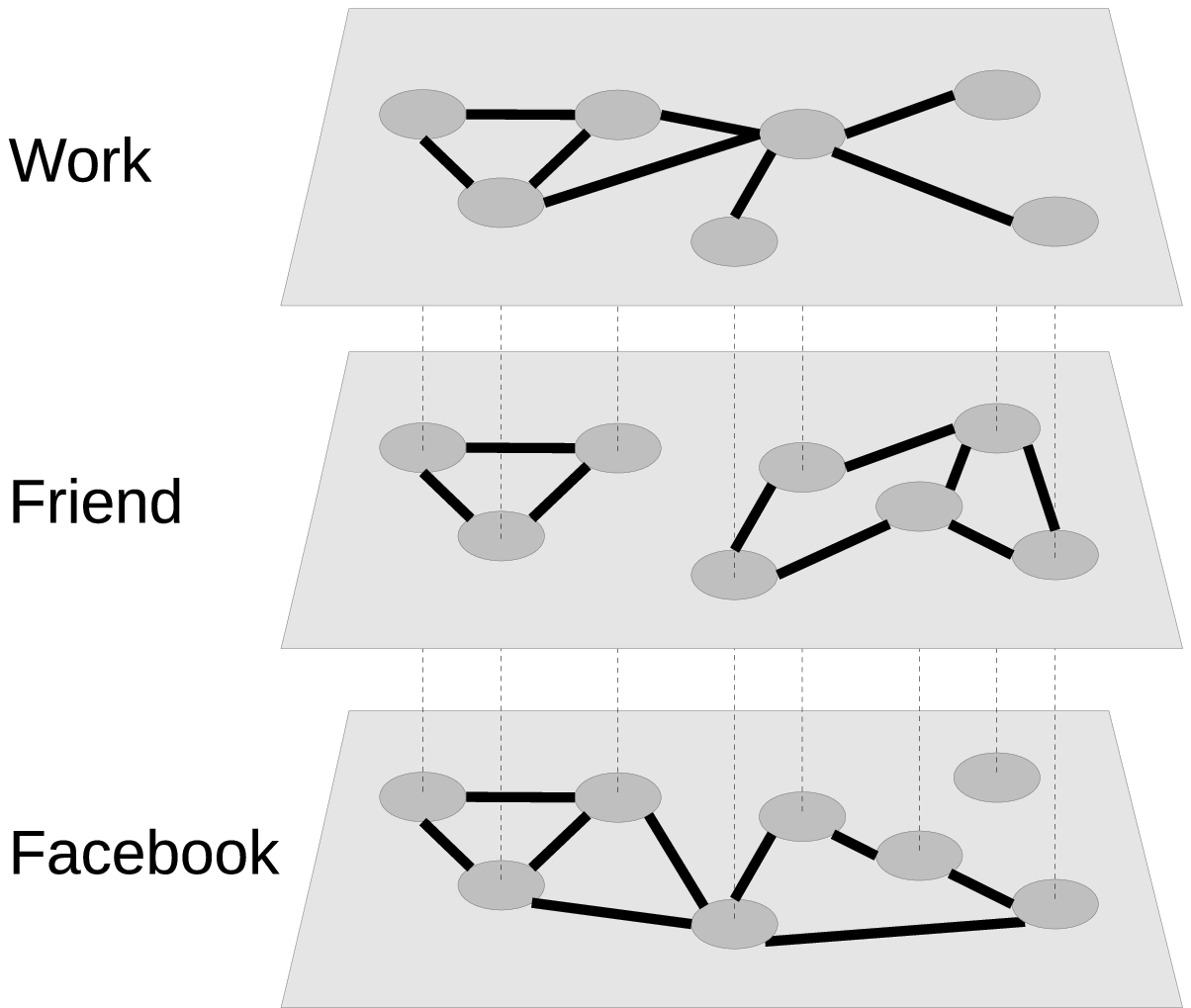}
    \label{fig:multilayer}
}
\subfigure[]{
    \includegraphics[width=.25\textwidth]{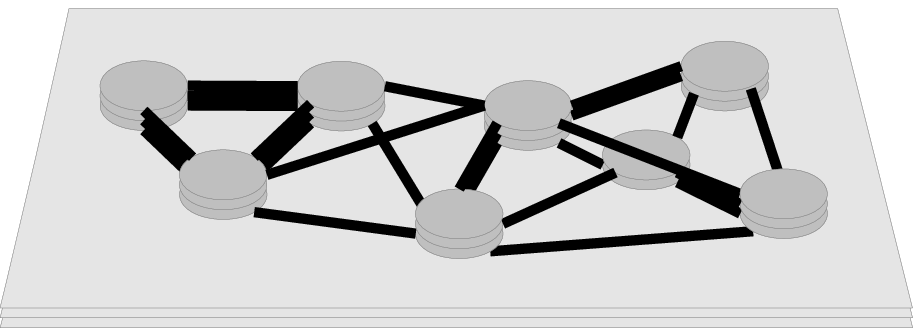}
    \label{fig:flattening}
}
\subfigure[]{
    \includegraphics[width=.25\textwidth]{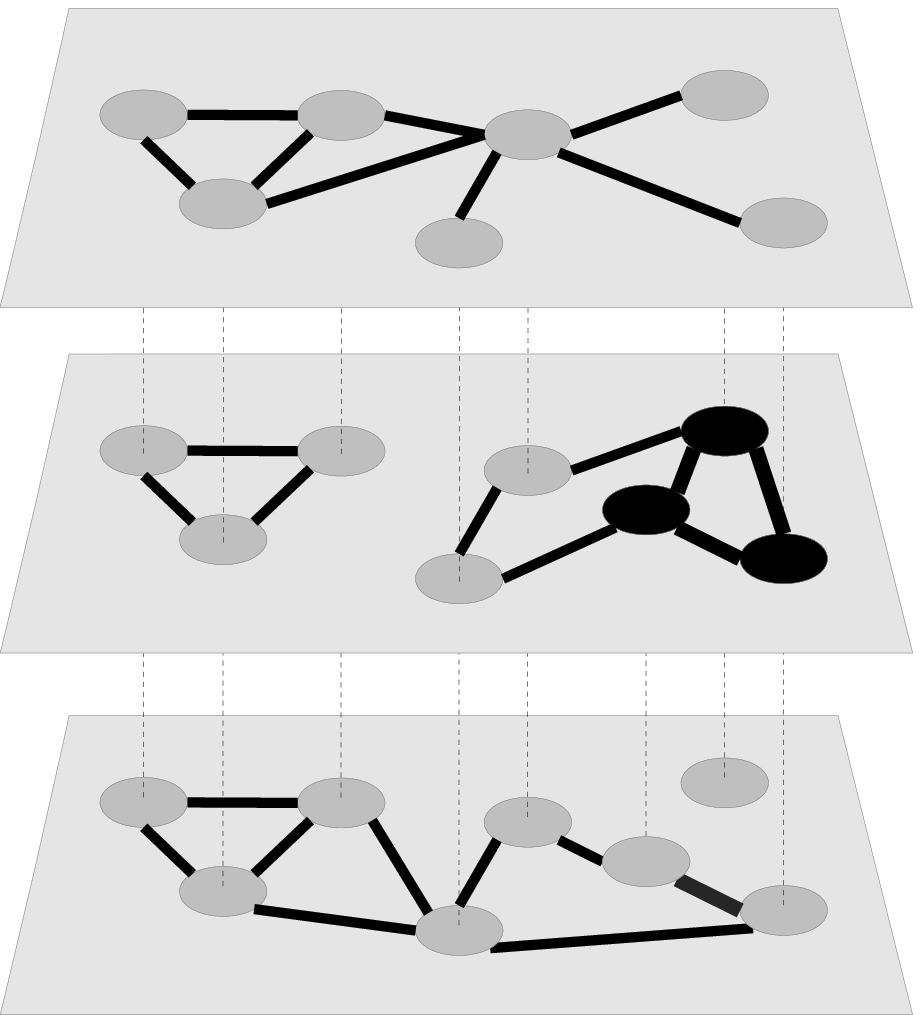}
    \label{fig:exclusive}
}
\caption{An edge-attributed graph, corresponding to a set on interconnected graphs defined on a common superset of individuals \subref{fig:multilayer}. An indirect way to process it is to reduce it to a single weigh\hlt{t}ed graph, then apply classical clustering algorithms \subref{fig:flattening}. A significantly different approach is to look at exclusive connections \subref{fig:exclusive}}
\label{fig:basic_model}
\end{figure}

\hlt{Although very similar, and in this specific example equivalent, these two representations emphasize different aspects of an edge-attributed graph. It is important to understand that the methods covered in the remaining of this section have been developed starting from specific models, influencing their features. R}esearchers using the first model have mainly focused on the reduction of different edge types to single edge\hlt{s, while r}esearchers using the second model have looked for clusters spanning different \hlt{layers and nodes belonging to multiple clusters depending on the edge type}. With this difference in mind, in the following we will formally represent both scenarios using the second (more general) model, where a family of graphs possibly containing common nodes represent the different kinds of edges. A larger working example is shown in Figure~\ref{fig:basic_model}\subref{fig:multilayer}.

More general definitions have been provided in the literature, where one node in one graph can correspond to multiple nodes in another. This includes the case of online social media, where the same user can open multiple accounts on some services \citep{MagnaniASONAM2011}, and the case of non-social networks containing different kinds of nodes, such as a power grid and a control network, where one node in a network can be related to multiple nodes in another \citep{Gao2011a}. \hlt{Similarly, the model introduced by }\hlc{\citet{Kivela2013}}\hlt{ allows the presence of attributes both on nodes and edges.} For the sake of simplicity we focus on the simpler definition\hlt{s} above\hlt{,} because \hlt{they are} the one\hlt{s} used by almost all works on \hlt{clustering} social networks to date. Also, notice that we focus on nominal attributes, e.g. \emph{work} and \emph{friendship}: the case where attributes are only numeric, that is, weigh\hlt{t}ed graphs, has already been treated in depth in existing surveys. However, we will deal with numeric weights when these are used inside algorithms for nominal attributes.

\subsection{Single-layer approaches}

A basic approach to deal with edge-attributed graphs is to \emph{flatten} them: to reconstruct a single weigh\hlt{t}ed graph so that existing clustering methods can be indirectly applied. This approach, exemplified in Figure~\ref{fig:basic_model}\subref{fig:flattening}, is not restricted to clustering but can be applied to any operation defined on weigh\hlt{t}ed graphs. Weights can be computed straightforwardly so that an edge between two nodes has a weight proportional to the number of graphs where the two \hlt{nodes} are \hlt{directly} connected. 
\begin{definition}[Flattening]\label{def:flattening}
A flattening of an edge-attributed graph $(\{G_i\})$ is a weighted graph $(E_f,V_f,w_f)$ where $E_f = \bigcup E_i, V_f = \bigcup V_i$ and $w(u,v) = \frac{|\{i\ |\ (u,v) \in E_i\}|}{N}$ (where $N$ is the total number of graphs).
\end{definition}
\citet{Berlingerio2011a} follows this approach. However, the same authors point out how this solution may discard relevant information, e.g., the fact that some attribute values (or graph layers) are more important than others to define a cluster.
\citet{Tang2011} \hlt{propose a more general framework where the} information about the multiple \hlt{edge types is considered during} one of the four different components of the community detection process, network flattening being one of them. 
Nevertheless, the authors point out that this kind of integration requires that \hlt{edges of different types} share the same community structure. Therefore, it is not suitable for cases where the structures significantly vary in \hlt{different} dimensions.

An antithetic approach acknowledging the importance of edge-attributed models but still not considering clusters that can span several graphs is introduced by \citet{Bonchi2012}. While flattening tends to assign \hlt{nodes directly} connected on multiple graphs to the same group because they get connected by a strong edge in the flattened graph, \citet{Bonchi2012} consider a set of \hlt{nodes} as a good cluster if their relationships are as specific and homogeneous as possible, i.e., they \hlt{are mainly connected through the same edge type}. An example is presented in Figure~\ref{fig:basic_model}\subref{fig:exclusive} where the three nodes marked in black are  connected \hlt{with each other} in the middle \hlt{layer but only share one single edge} on all other \hlt{layers}, representing a good cluster according to this approach\footnote{Please notice that this specific example is not compatible with the original model by \citet{Bonchi2012} where individuals are allowed to be \hlt{directly} connected only on one of the \hlt{layers}. However, it retains its underlying intuition. \hlt{While this work was not originally intended to be applied to this domain, it still presents a worth-mentioning alternative point of view.}}.

\hlt{The n}ext sections are devoted to methods aiming at identifying clusters \hlt{spanning} multiple \hlt{layers}. They are mostly extensions of quality measures \hlt{traditionally} used in graph clustering, modularity and quasi-cliques being two prominent \hlt{examples}.

\subsection{Extension of modularity} \label{modularity-based}

\begin{figure}[ht]
\centering
\includegraphics[width=.65\textwidth]{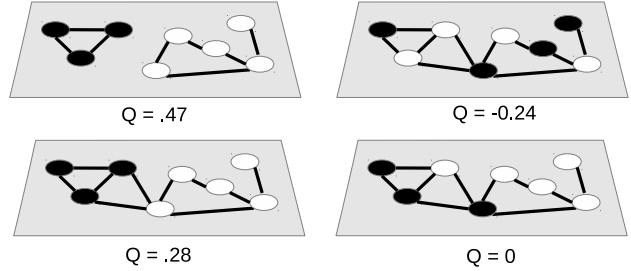}
\caption{Modularity of four graph clusterings: nodes in each graph are assigned to two clusters (black and white); the modularity of each assignment is reported under the graph}
\label{fig:modularity}
\end{figure}

Modularity is a measure of how well the nodes in a graph can be separated into dense and independent components \citep{Newman2004}. Figure~\ref{fig:modularity} shows four graphs with their nodes assigned into two communities (black and white) and the modularities resulting from these assignments. In these examples it clearly appears how the assignments putting together highly interconnected nodes and separating groups of nodes with only a few connections between them get a higher value of modularity. It is worth noticing that modularity is not a method to find communities, but only a quality function. However, it can be directly optimized or used inside community detection methods to guide the clustering process.

Although this measure suffers from some well known pitfalls \citep{Fortunato2007,AndreaLancichinettiSantoFortunato2011}, it has recently been at the basis of several graph clustering methods and it has also been extended to deal with attributed graphs. Let us briefly introduce it\footnote{Please notice that modifications of this formula have been proposed to make it more adaptable to different datasets. One typical addition is a \emph{resolution parameter}, that we have omitted from the following equations because it is orthogonal to our discussion.}, to later simplify the explanation of its extension. \hlt{The modularity is thus expressed as}

\begin{equation}\label{eq:modularity}
Q = \frac{1}{2m} \sum_{ij} \left(a_{ij} - \frac{k_i k_j}{2m}\right) \delta(\gamma_i,\gamma_j),
\end{equation}

\hlt{where $\delta(\gamma_i,\gamma_j)$ is the Kronecker delta which} returns $1$ when nodes $i$ and $j$ belong to the same cluster, $0$ otherwise. Therefore, the sum is computed only for those pairs of nodes that are inside the same cluster. For each of these pairs, \hlt{the presence of an edge between them} improves the quality of the assignment: $a_{ij}$ equals 1 when \hlt{there is an edge between} $i$ and $j$, 0 otherwise. As we are dividing everything by $m$ (the number of edges in the graph), \hlt{edges between} nodes \hlt{belonging to} different clusters negatively affect modularity because \hlt{they are} not considered in the numerator (as $\delta(\gamma_i,\gamma_j)=0$), but \hlt{are} counted in the denominator ($m$). Finally, the formula considers the fact that two nodes with high degree would be more likely to end up in the same cluster by chance, therefore their contribution is reduced ($- \frac{k_i k_j}{2m}$, where $k_i$ and $k_j$ are the degrees of $i$ and $j$).

\begin{figure}[ht]
\centering
\includegraphics[width=.35\textwidth]{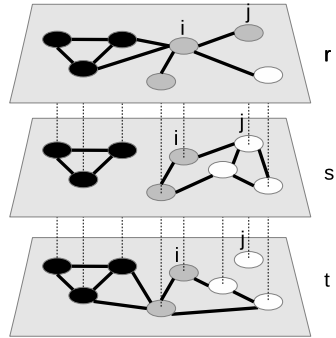}
\caption{An edge-attributed graph with three kinds of edges, represented as three interconnected graphs. Nodes have been assigned to three clusters (black, gray and white)}
\label{fig:mucha}
\end{figure}

Now it should be easier to understand the extension of modularity proposed by \citet{Mucha2010a} for edge-attributed graphs. Let us consider Figure~\ref{fig:mucha}: here we have emphasized how the same individual $i$ can be present in multiple graphs at the same time. For example, $i$ and $j$ are \hlt{directly} connected on graphs $r$ and $s$, where $r$ and $s$ represent two different edge types. Notice that in this example we have three graphs, i.e., three edge types, and that $j$ is assigned to two different clusters in graphs $r$ (gray) and $s, t$ (white). 

\hlt{Thus, the extended version of the modularity can be expressed as}
\begin{equation}
Q_m = \frac{1}{2 \mu} \sum_{ijsr} \left[\left(a_{ijs} - \frac{k_{is} k_{js}}{2m_s}\right)\delta(s,r) + c_{jsr} \delta(i,j)\right] \delta(\gamma_{i,s},\gamma_{j,r}).
\end{equation}

This extended quality function involves not just all pairs of nodes ($i, j$) but also all pairs of graphs ($s, r$).
$\mu$ and $\delta(\gamma_{i,s},\gamma_{j,r})$ correspond respectively to $m$ and $\delta(\gamma_i,\gamma_j)$ in the modularity formula, where $\mu$ also considers the \hlt{connections} between different graphs\hlt{: we say that there is a connection between two} graphs $r$ and $s$ \hlt{whenever they contain a common node $j$, } \hlt{which increases} $\mu$ by $c_{jsr}$\hlt{.} $\delta(\gamma_{i,s},\gamma_{j,r})$ allows to assign the same \hlt{node} to different clusters inside different graphs. The sum is now made of two components. One is only computed when two nodes in the same graph are considered (because of $\delta(s,r)$), corresponding to modularity. In fact, here $a_{ijs}=1$ when $i$ and $j$ are \hlt{directly} connected in graph $s$ and $k_{is}$ is the degree of node $i$ in the same graph. The second component, $c_{jsr}$, is only computed when we are considering the same node $j$ inside two different graphs $r$ and $s$. This term increases the quality function by $c_{jsr}$ (typically, a constant value ranging from 0 to 1) whenever we assign the same individual to the same cluster on different graphs.

One practical problem in using this measure is to set the $c_{jsr}$ parameter. Setting it to 0 for all nodes and graphs, clusters are identified on each single graph independently of each other. If $c_{jsr}$ is high, e.g., 1, it becomes unlikely to assign the same individuals to different clusters on different graphs. Other practical aspects to consider are the fact that the part of the formula corresponding to traditional modularity can give a negative contribution, which is not true for the part taking care of inter-network relationships, and also the fact that the contribution of inter-network relationships grows quadratically on the number of networks while the modularity part only grows linearly. However, while the choice of appropriate parameters deserves more research, this extended definition of modularity can be directly used to find clusters by using any modularity-optimization heuristics, as done by \citet{Mucha2010a}, or paired with a concept of betweenness to extend the Girvan-Newman algorithm. The definition of betweenness for edge-attributed graphs follows directly from any definition of distance involving multiple graphs \citep{Brodka2011,Magnani2013a}.

\begin{figure}[ht]
\centering
\includegraphics[width=.65\textwidth]{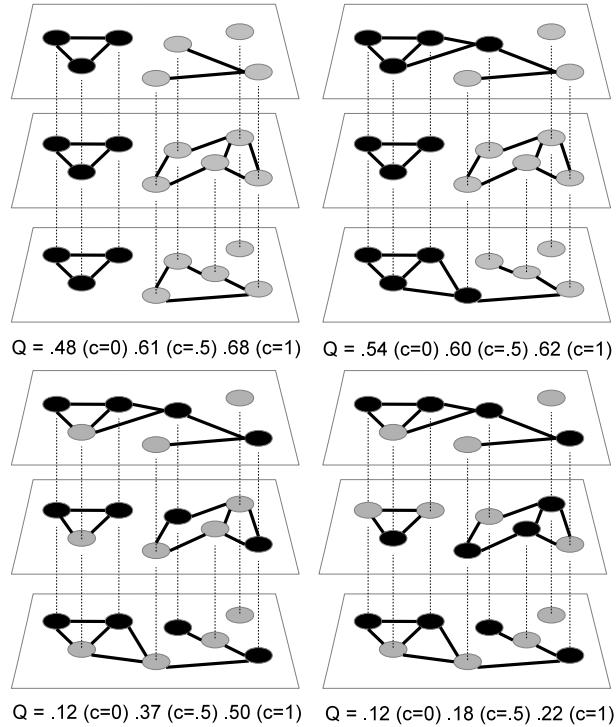}
\caption{Multi-layer modularity of four graph clusterings: nodes in each graph are assigned to two clusters (black and gray); the modularity of each assignment is reported under the graph using three settings: $c_{jsr}=0$, $c_{jsr}=.5$ and $c_{jsr}=1$}
\label{fig:mucholarity}
\end{figure}

Figure \ref{fig:mucholarity} shows the values of modularity for four different multi-graphs and three different settings for the inter-graph parameter $c_{jsr}$ (which is kept constant for all nodes and graphs). The figure emphasizes the different components of this measure. On the top we can see two clusterings aligned with both the single-graph and mu\hlt{lt}i-graph structure. In particular, groups of nodes \hlt{sharing several edges} belong to the same cluster, and the same nodes on different graphs tend to belong to the same cluster. However, the top-right example shows that we can assign a node to different clusters in different graphs.

Modularities computed using different values of $c_{jsr}$ cannot be compared: increasing $c_{jsr}$ also increases the absolute value of modularity. However, we can see how the increase in the top-right figure is proportionally lower than the one on the left (from .48 to .68 and from .54 to .62, respectively). This is determined by the nodes assigned to multiple clusters.

The two lower figures show examples of lower modularity, i.e., clusterings not following the structure of the graphs. The lower-left image has a low overall intra-graph modularity which can be seen when $c_{jsr}=0$ and thus inter-graph connections are not considered. When we also consider them ($c_{jsr} = .5$ and $c_{jsr} = 1$) we can see that modularity is increasing in the lower-left graph much more than in the lower-right one, where every node belongs to both clusters on different layers.

\subsection{Clique-finding methods}

Another concept used to discover clusters in graphs is the clique, i.e., a complete (sub)graph. Although this is one of the basic concepts in graph theory and it is thus well known, we briefly recall it.
\begin{definition}[Clique]
A clique is a set of nodes \hlt{directly} connected to all other nodes in the clique.
\end{definition}

\begin{figure}[ht]
\centering
\subfigure[]{
    \includegraphics[width=.17\textwidth]{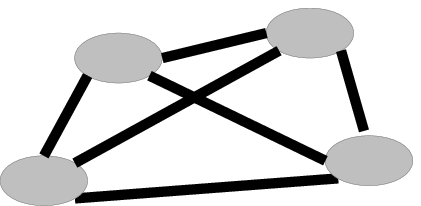}
    \label{fig:clique}
}
\subfigure[]{
    \includegraphics[width=.17\textwidth]{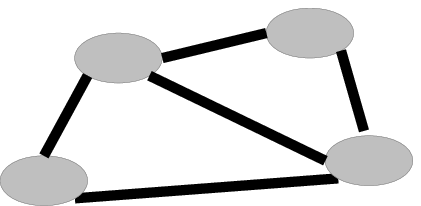}
    \label{fig:quasiclique}
}
\subfigure[]{
    \includegraphics[width=.17\textwidth]{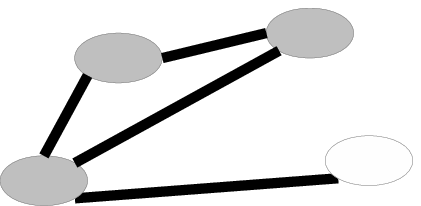}
    \label{fig:nonclique}
}
\caption{A clique \subref{fig:clique}, a quasi-clique \subref{fig:quasiclique} and four nodes not making a .5-quasi-clique \subref{fig:nonclique}}
\label{fig:cliques}
\end{figure}

\begin{definition}[Maximal clique]
A maximal clique is a clique that is not contained in a larger clique.
\end{definition}
Figure~\ref{fig:clique} shows an example of a clique. Any three nodes in Figure~\ref{fig:clique} still make a clique, but not a maximal one because we can add the fourth node and still have a clique.

A (maximal) clique clearly corresponds to a cluster. However, large cliques are difficult to find in real data because it is sufficient for one edge not to be present to break the clique, and in social graphs edges can be missing for many reasons, e.g., because of unreported data or just because even in a tight group there can be two individuals that do not get well together. Therefore, when clustering is applied to social graphs, it is wiser to look for more relaxed structures called quasi-cliques.

\hlt{For example,  }\hlc{\citet{Freeman1996}}\hlt{ studies the cliques gathered from interviews to a group of individuals
and acknowledges that they are not enough for defining communities.
}

\begin{definition}[Quasi-clique]
A quasi-clique is a set of nodes where each node is \hlt{directly} connected to at least $\gamma \%$  of the other nodes in the quasi-clique.
\end{definition}

Algorithms to discover quasi-cliques take $\gamma$ as a parameter. Please notice that similar alternative definitions are possible, e.g., using a strict $>$ or considering the percentage over all nodes in the quasi-clique --- the underlying concept remains the same. In Figure~\ref{fig:quasiclique}, we have illustrated a .5-quasi-clique, and in Figure~\ref{fig:nonclique}, we have four nodes that do not constitute a .5-quasi-clique because the white node is \hlt{directly} connected to only one third of the other nodes.

The problem of finding quasi-cliques in a graph \hlt{is} NP-hard. \hlt{According to common beliefs, this implies that no algorithm can exactly solve this problem in a reasonable amount of time even for small graphs. However, efficient algorithms which do not guarantee the identification of all quasi-cliques have been proposed}.

As previously mentioned, the most common interpretation of clusters in edge-attributed graphs states that multiple kinds of edges between two individuals strengthen their relationship. Therefore, \citet{Pei2005} have introduced algorithms to discover quasi-cliques in all graphs and \citet{Wang2006,ZhipingZeng2006} to identify quasi-cliques in at least a given percentage of graphs (where this threshold is called \emph{support}).

While not based on quasi-cliques, the ABACUS algorithm by \citet{Berlingerio2013} also applies a similar definition, coming from the \emph{frequent itemset mining} problem. First, clusters are identified in each graph, then those individuals being in the same cluster in at least a given percentage of graphs are also included into a global cluster in the final result.

It is worth noticing that quasi-clique clustering methods were first developed for generic graph databases without focusing on the application domain of social graphs. In this specific domain, while we may agree that a cluster spanning all the graphs represents a strong global cluster, a group of \hlt{nodes sharing a large number of edges} on a few specific graphs may also identify a cluster of interest. For example, we might find that a group of individuals goes to the same school and plays in the same basketball team. This is a strong relationship that should not be negatively affected by the existence of other relationships where they do not form a group. However, adding other edge types to the attributed graph (which corresponds to adding new graphs to the multi-layer graph structure) would reduce their support.

The approach proposed by \citet{Boden2012} starts from this consideration and looks for sets of nodes that make a cluster in each single graph of \emph{any subset} of the graphs in an edge-attributed model. This work also considers the case of weigh\hlt{t}ed graphs, but this is peculiar to this method and we will not provide additional details here.

\subsection{Emerging clusters} \label{emerging-clusters}

We conclude this section presenting a hypothesis still unverified in the literature that in our opinion might lead to the development of new clustering methods. The hypothesis is that clusters can emerge when a specific combination of graphs is considered, and disappear when more graphs are added to the model. 

\begin{figure}
\centering
\includegraphics[width=.8\textwidth]{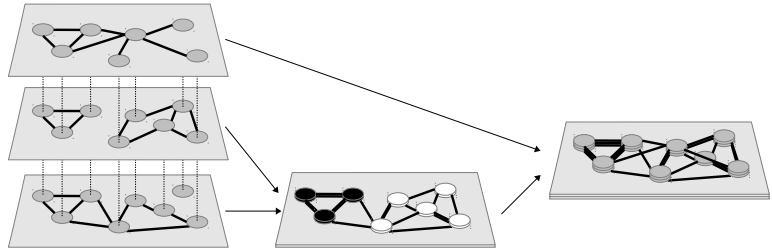}
\caption{Emerging clusters: well separated clusters appear when a specific subset of the graphs is used, but disappear when less or more networks are added}
\label{fig:emerging}
\end{figure}

In Figure~\ref{fig:emerging}, the idea is illustrated on a simple example. The analysis of the three graphs together (right hand side of the figure) does not reveal any interesting patterns as there are too many edges in the graph. The same can be observed for each single graph (on the left). However, choosing two specific layers, some more evident clusters emerge (center, clusters denoted by black and white nodes). None of the previously presented approaches seems to be able to find such clusterings, because they require every cluster to be present in at least one of the single graphs.

This hypothesis would also provide an answer to the difficulty in finding good clusterings in real social graphs. In fact, although several clustering algorithms exist, in practice they achieve good results when some more or less well-separated clusters exist. This is strictly related to the way in which community detection algorithms have been defined: some try to maximize modularity, favoring well separated clusters, some use random walk approaches, where the probability that a walker crosses two clusters is proportional to the number of edges between them, some exploit measures like betweenness, that is high when few other edges connect distinct portions of the graph \citep{Fortunato2010}. However, when we deal with on-line relationships, clustering becomes extremely hard. According to our hypothesis, this depends on the fact that a large number of semantically different layers are considered all-together, determining the co-existence of several overlapping clusters, and a case of information overload. 

In summary, if we consider Figure~\ref{fig:emerging} (right side), we would not expect any clustering algorithm to find evident clusters. However, in theory clusters may appear when the multi-layer \hlt{organization} of the edges is unfolded in specific ways, e.g., by only retaining the two layers in Figure~\ref{fig:emerging} (center). Therefore, the problem shifts from being purely algorithmic (e.g., how do we find the best cut?) toward aspects like the choice of the data model, data preprocessing and feature selection.

A preliminary work in this direction that can be seen as a conjunction between the idea of emerging clusters and the flattening approach is discussed by \citet{Rocklin2011}.
This work proposes an algorithm to find a vector that weights the layers to aggregate them such that the clustering of the resulting flattened graph is as similar to a given ground-truth clustering as possible (the clustering algorithm and a similarity measure between weighted single-layer graphs are given for this problem).
The second half of the paper deals with the rich clustering structure that the multi-typed edges can provide.
Generating random aggregates of the graph, the authors explore the space of possible clusterings and study, e.g., if good graph clusterings are \emph{clustered} in this space. The final problem that they tackle is how to give an efficient representation of this resulting meta-clustering. Their approach is to reduce each meta-cluster (of clusterings) into a single representative clustering and select a small number of them to cover the meta-clustering space. In this way, they provide a set of diverse and non-redundant clusterings as output.

\section{Clustering node-attributed graphs}\label{sec:nodes}


According to the 
taxonomy presented by \citet{Getoor2005Survey}, node-attributed graph clustering aims at detecting groups of nodes sharing common characteristics considering both their attributes and their position in the graph. Most of the works addressing this problem are based on \emph{partitioning} and \emph{homophily}: nodes can belong to one and only one group, and nodes in the same group must have homogeneous values on their attributes. A few other methods, also covered here, generate overlapping clusters, e.g., by considering different combinations of the attributes. This last approach is usually known as \emph{subspace} clustering.

\subsection{Data representation}


Like in the case of edge attributes, also when attributes on nodes are considered, the literature abounds with terminologies and models depending on the research field or the finality of the work, making it difficult to provide a unified view. However, we can see some main options emerging.

As previously mentioned, \citet{Wasserman1994} describe multiple dimensions that can be represented in a social network model: a \textit{structural} dimension (relationships among actors), a \textit{compositional} dimension (attributes of the single actors), and an \textit{affiliation} dimension (representing group memberships). Affiliation information often refers to known groups such as clubs or companies, but it can also represent the cluster memberships discovered through a clustering process.

\begin{figure}[ht]
\centering
\subfigure[]{
    \includegraphics[width=.48\textwidth]{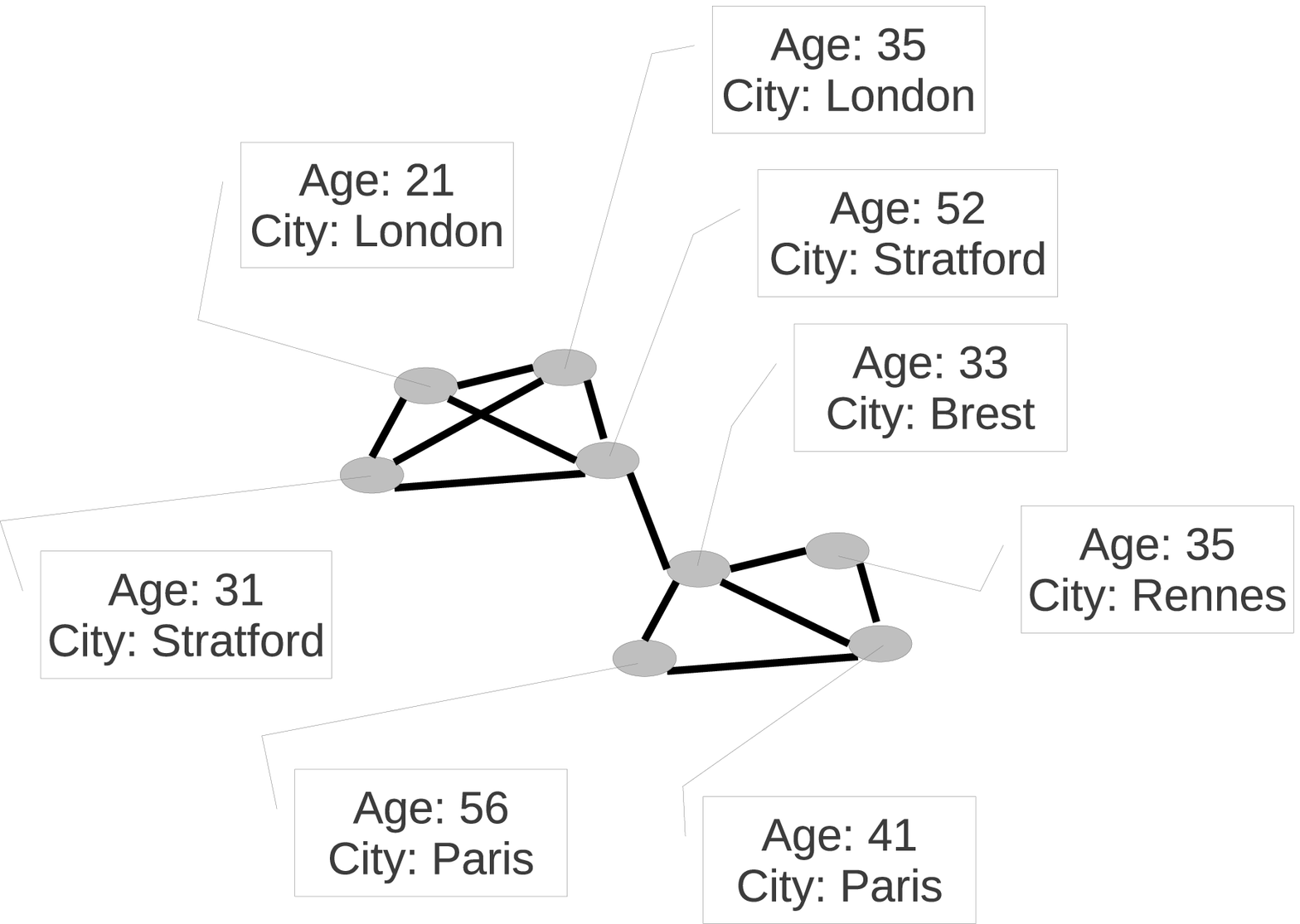}
    \label{fig:attributed_graph_2}
}
\subfigure[]{
    \includegraphics[width=.48\textwidth]{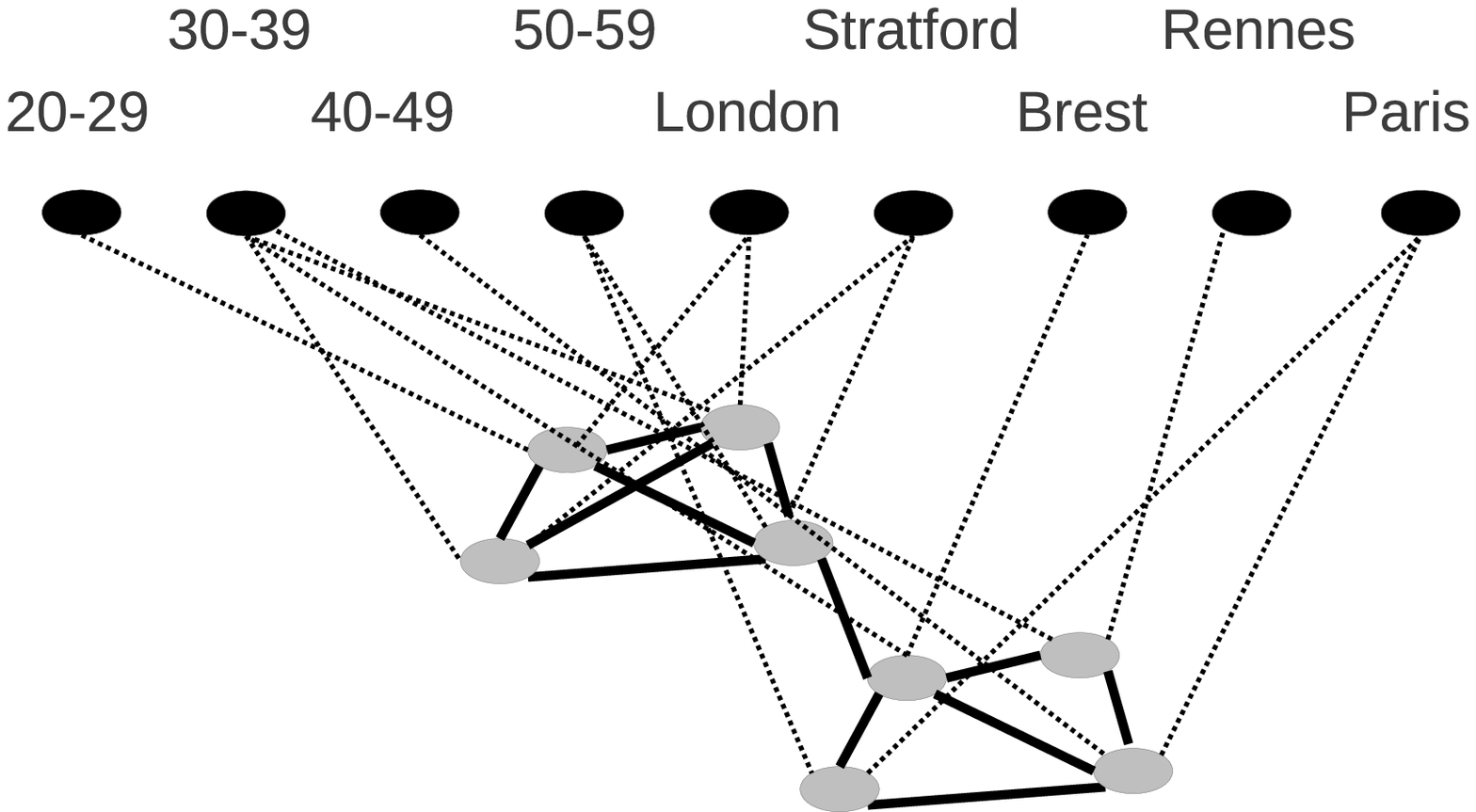}
    \label{fig:attribute_augmented_graph}
}
\caption{\subref{fig:attributed_graph_2} Attributes represented as tuples describe node properties. 
The similarity/distance between tuples can be integrated into the graph and used during the 
clustering process. \subref{fig:attribute_augmented_graph} New nodes representing the additional information are added to the original graph,
resulting in a heterogeneous structure with multiple node types.}
\label{fig:node_attribute_representation}
\end{figure}

Two main options to represent such a model are shown in Figure~\ref{fig:node_attribute_representation}. The first one, Figure \ref{fig:attributed_graph_2}, consists in extending a structural graph  
with tuples describing node properties. This can be formally expressed as a triple $G=\left(V,E,F\right)$ where each node $v$ is associated with a set of $a$ attributes (or a \emph{feature vector}) $[f_1(v), ... f_a(v)]$, storing its compositional dimension. Note here that the affiliation information may be stored in the same way, by adding attributes dedicated to memberships. The second option, Figure \ref{fig:attribute_augmented_graph}, consists in superimposing one or more graphs where additional nodes represent either specific attribute values or groups. Structurally, this superimposed graph is \emph{bipartite} because it connects individuals to groups, without edges between groups or between users (the latter are stored in the original social network). More formally, a graph $G_p=\left(V_p,E_p\right)$ is augmented by a bipartite graph $G_a=(V_p \cup V_a,E_a)$, connecting nodes of $V_p$ to attribute nodes of $V_a$, with no links between attributes: $E_a \subseteq V_p \times V_a$. This defines an augmented graph $G=\left(V,E\right)$ with $E=E_p \cup E_a$ and $V=V_p \cup V_a$.  

Several terms have been used in the literature to refer to the options presented in Figures \ref{fig:attributed_graph_2} and \ref{fig:attribute_augmented_graph}, or even for their intermediate variations. To make access to the existing literature easier, in Table~\ref{node-attributed-terminology} we report the main terms together with the references to where they appear and the indication of which modeling option has been adopted.
Our objective here is not to be exhaustive: we aim at capturing the relationships between different approaches.
%
For example when \citet{Tong:2007:FDP:1281192.1281272} refer to an \emph{attribute graph}, they imply that they have previously grouped the nodes with common attributes, and propose a meta-graph where meta-nodes reflect those groups and edge weights represent group-to-group similarity. 
\citet{zheleva:kdd09} study \emph{social and affiliation networks} keeping two distinct graphs and observing the co-evolution of these two graphs via their common nodes, retrieved from Flickr groups.
In the machine learning field, in the late 1990s and early 2000s, workshops  dedicated to \textit{link mining} referred to \textit{relational data} \citep{Neville2003}. In a more recent data warehousing context, \citet{zhao2011graph} introduced an OLAP graph cube for \emph{multidimensional networks}.

\begin{table}
\caption{\hlt{Some terminology used in the literature to refer to node-attributed graphs}}
\label{node-attributed-terminology}
{\setlength{\tabcolsep}{.1cm}
\begin{tabular}{lm{5cm}l}
\hline 
\textbf{term} & \textbf{references} & \textbf{option} \\ 
\hline 
Social-attribute network & \citep{Yin:2010:LUF:1772690.1772879, Yin:2010:SAN} & (b)\\
\hline 
Attribute augmented graph & \citep{Zhou2009, Zhou2010}  & (b) \\
\hline 
Attributed graph & \citep{Zhou2009, Cruz2013a, Cruz2013b}  & (a) \\
\hline 
Feature-vector graph & \multirow{2}{*}{\citep{gunnemann2013efficient}} & \multirow{2}{*}{(a)} \\
Vertex-labeled graph &  & \\
\hline
\end{tabular}}
\end{table}

In summary, there has not been a consensus on the model yet. While different formats are useful to emphasize different aspects, all models include both structural and compositional data and one can be derived from another. Therefore, to introduce existing methods, we will use a common model consisting of an attributed graph $G=\left(V,E,F\right)$ where nodes are associated with an attribute vector $F(v)$. 


\subsection{Weight modification according to node attributes} \label{weights-modification}

The first class of methods we present is based on the following idea: first the node-attributed graph is reduced to a single weighted graph, where weights represent
attribute similarity. Then, any clustering algorithm for weigh\hlt{t}ed graphs can be applied in principle. Different methods use alternative functions to compute node similarity and to update edge weights when similarities have been computed. However, in all these approaches the change of weights influences the clustering algorithm to privilege the creation of groups in which the nodes are not only well connected but also similar. 

\begin{figure}
\centering
\subfigure[]{
    \includegraphics[width=.48\textwidth]{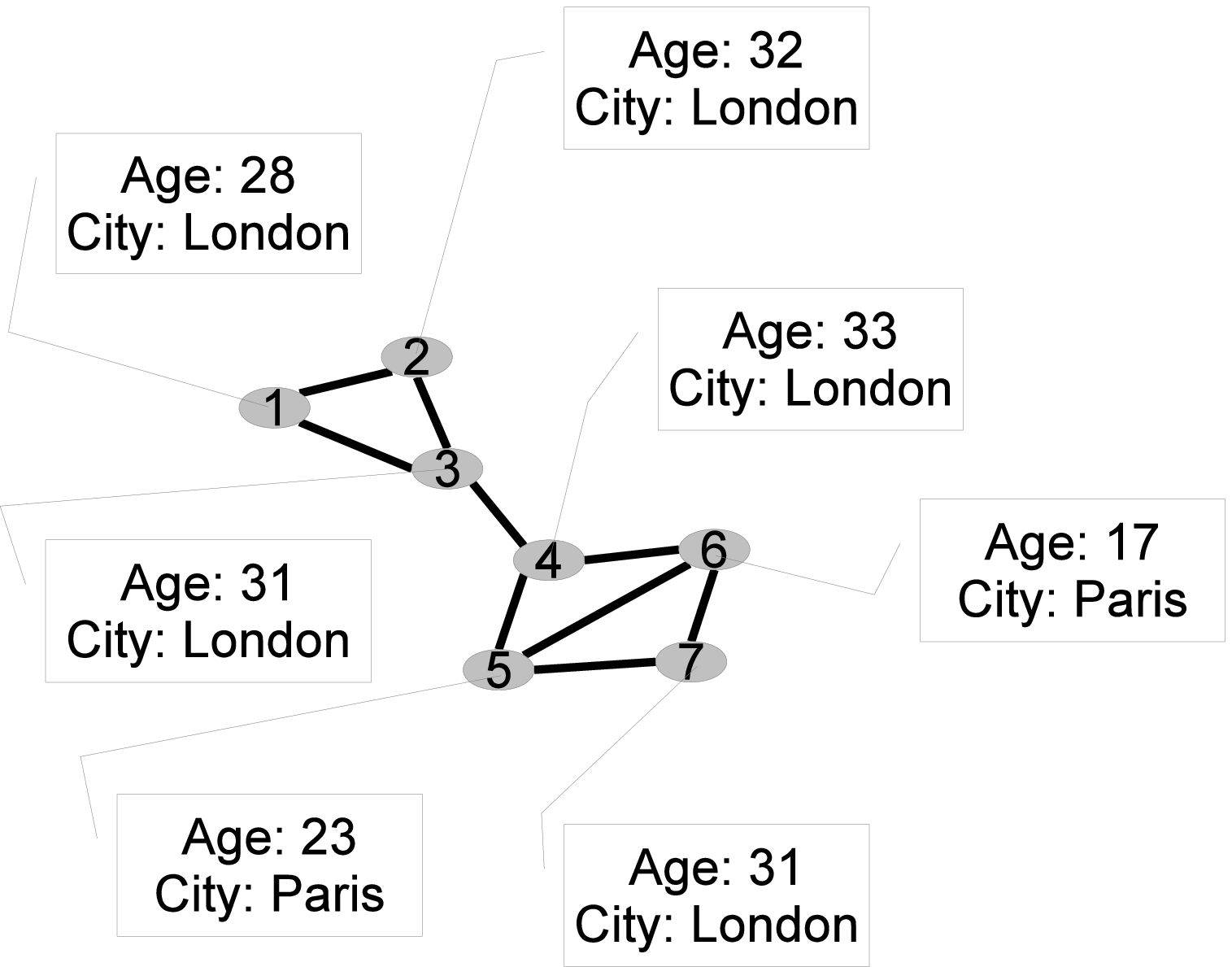}
    \label{fig:node-attribtue-methods}
}
\subfigure[]{
    \includegraphics[width=.28\textwidth]{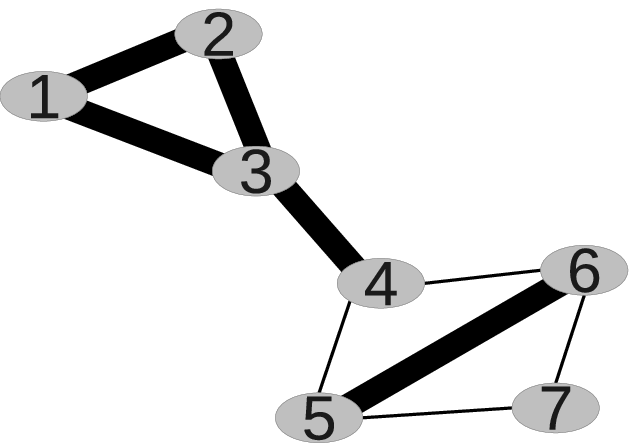}
    \label{fig:node-attribute-weights}
}
\caption{A node-attributed graph (a) and an attribute-free representation of the same graph (b) where attribute similarities are stored in the edge weights (b). Thicker edges indicate a higher weight, i.e., a stronger connection}
\label{fig:node-attribute-methods}
\end{figure}

As an example, consider Figure \ref{fig:node-attribute-methods}. Focusing solely on the attributes, nodes $\{1, 2, 3, 4, 7\}$ would form a homogeneous cluster, well separated from nodes $\{5, 6\}$. If we only consider the structure of the graph, two clear clusters emerge (nodes $\{1, 2, 3\}$ and nodes $\{4, 5, 6, 7\}$). These two pieces of information are summarized in the weighed graph in (b). While the specific final clusters depend on the assigned weights, we can see the emergence of a cluster made of nodes $\{1, 2, 3, 4\}$, presenting both structural and compositional similarities and otherwise difficult to identify.
Table~\ref{tab:weight-update} summarizes the main works adopting this strategy, and the measures mentioned in the table are reported in the following.

\begin{table}
\caption{\hlt{Variations of the \emph{weight modification} approach}}
\label{tab:weight-update}
{\setlength{\tabcolsep}{.1cm}
\begin{tabular}{m{4cm} m{4cm} m{4cm}}
\hline 
\textbf{reference} & \textbf{similarity} & \textbf{clustering} \\ 
\hline 
\multirow{3}{*}{\citep{Neville2003}} & \multirow{3}{*}{Matching coefficient}  & Karger's Min-Cut \\
 & & MajorClust\\  
 & & Spectral\\  
\hline
\citep{Steinhaeuser2008} &  Extended matching coefficient & Assign $u$ and $v$ to the same cluster when  the weight of ($u$,$v$) is above a given threshold \\
\hline
\citep{Cruz2011a} &  \multirow{2}{*}{Self-organizing maps} & \multirow{2}{*}{Louvain} \\
\citep{Cruz2012} & & \\
\hline
\end{tabular}}
\end{table}

For example, \citet{Neville2003} use the \textit{matching coefficient} similarity metric $S_{ij}$ quantifying the number of attribute values ($k$) the nodes have in common. \hlt{This similarity metric is expressed as}
\begin{equation}\label{eq:single_matching_coeff}
S_{ij}=\begin{cases}
\sum_{k} s_{k}\left(i,j\right) & \text{ if } e_{ij}\in E  \text { or } e_{ji}\in E \\
0 & \text{otherwise} \\
\end{cases},
\end{equation}
where 
$$
s_{k}\left(i,j\right)=\begin{cases}
1 & \text{ if } k_{i} = k_{j} \\
0 & \text{otherwise} \\
\end{cases}.
$$

Once the weights have been changed, the graph is clustered using one of the three methods reported in Table~\ref{tab:weight-update}: Karger's Min-Cut \citep{Karger:1993:GMR:313559.313605}, MajorClust \citep{stein1999nature} or spectral clustering with a normalized cut objective function \citep{shi2000normalized}. Experimenting with artificial datasets, spectral clustering appears to be robust to irrelevant attributes and graphs with low linkage.


\citet{Steinhaeuser2008} extend the matching coefficient computation to take both discrete and continuous attributes into account: for discrete attributes, each common attribute shared by two nodes increments the weight of $e\left(u,v\right)$ by $1$; for continuous attributes, 
the idea is to add the normalized distance between the attributes.
Once the weights have been changed and normalized, all nodes, connected by an edge whose
weight is greater than a threshold $t$, are assigned to the same cluster. In this specific work the quality of the final partition is evaluated using
modularity \citep{Newman2004}.

The approach presented by \citet{Cruz2011a,Cruz2012} deals with the fact that not all attributes
may be relevant to determine the similarity between nodes.
When too many attributes are involved in the computation of traditional distance functions, e.g. Euclidean distance, we lose the ability to
discriminate between different nodes. In fact, the so-called \emph{curse of dimensionality} materializes in that all distances tend to converge to the same value. In addition, some attributes may need to be combined/transformed to
become relevant. Therefore, the authors use a classical machine learning approach developed by \citet{Kohonen1997} and known as self-organizing map (SOM)\footnote{Self-organizing maps have been proposed as a learning approach
that is robust to noise and can map high dimensional data into low
dimensionality spaces, e.g. text.}, to find the latent information 
worth to establish the similarity between the nodes. 
An edge between two nodes from the same cluster gets its weight
strengthened proportionally to a given constant $\alpha \gg 1$.
The resulting weigh\hlt{t}ed graph is finally clustered using the Louvain method \citep{Blondel2008} and the overall complexity is linear $\mathcal{O}(n)+\mathcal{O}(fn)+\mathcal{O}(m)$, where $n$ is the number of nodes, $f$ the number of attributes or features and $m$ the number of edges.
Additionally, the authors introduce the notion of \textit{point of view}:
by manually selecting subsets of attributes, it becomes possible to analyze the social network from different perspectives. 

It is worth noticing that this family of techniques produces new edge weights according to node attributes. If the original social graph is also weighted the two kinds of weights must be combined is some way, e.g., by multiplying them.

\subsection{Linear combination of attributes and structural dimensions}



The previous family of methods removes node attributes by storing their information inside the edges of the graph. Some studies adopt an opposite approach consisting in the removal of the network: structural information is stored into a similarity (or a distance) function between nodes. After defining this function, classic distance-based clustering methods can be applied.
As an example, \citet{combe2012} define a distance between nodes which is given by
\begin{equation}\label{eq:combe_dist}
d_{TS}\left(i,j\right)=\alpha\cdot d_{T}\left(i,j\right)+\left(1-\alpha\right)d_{S}\left(i,j\right),
\end{equation}
where $ d_{T}\left(i,j\right)$ and $d_{S}\left(i,j\right)$ are the attribute and structural similarity, respectively, between nodes $i$ and $j$ and $0\leq\alpha\leq1$ is a weighting factor. 
The authors leave the choice of the clustering method open. 
Another similar distance function by \citet{Dang2012}, as listed in Table~\ref{tab:distance},
is used to build a k-nearest neighbor graph in order to find clusters using the Louvain method \citep{Blondel2008}.

The main feature of these approaches is that nodes which are structurally far from each other in the social graph can result to be close in case of similar attribute values. As a consequence, and depending on the distance-based clustering method, clusters may contain 
disconnected portions of the graph.
\citet{Hanisch01072002} experiment with a similar approach on biological networks and gene expression data. After the computation of the combined distance, they apply hierarchical clustering and a statistical measure to define the cutting point of the dendrogram. 

\begin{table}
\caption{Similarity or distance functions combining structural and compositional dimensions}
\label{tab:distance}
{\setlength{\tabcolsep}{.1cm}
\begin{tabular}{l l}
\hline 
\textbf{reference} & \textbf{similarity or distance} \\ 
\hline 
\citep{combe2012} & $\alpha\cdot d_{T}\left(i,j\right)+\left(1-\alpha\right)d_{S}\left(i,j\right)$ \\
\hline
\citep{villavialaneix2013} & $\alpha_{0}K_{0}\left(i,j\right)+\sum_{d}\alpha_{d}K_{d}\left(c^{d}_{i},c^{d}_{j}\right)$ \\
 \hline
\citep{Dang2012}  & $\alpha\cdot G_{i,j}+\left(1-\alpha\right)\cdot simA\left(i,j\right)$ \\
 \hline
\end{tabular}}
\end{table}


While \citet{villavialaneix2013} share a similar purpose using a weigh\hlt{t}ing parameter to balance their components, they rely on \textit{kernels} to map the original (multi-space) data into an (implicit and unique) Euclidean space where SOMs can be
used. In this case authors define a multi-kernel
similarity function to combine composition and structure
as indicated in Table~\ref{tab:distance}. 
$K_{0}\left(i,j\right)$ indicates the kernel measuring structural similarity, $c^{d}_{i}$ is the $d$th label of node $i$ and $\alpha_{d}$
are weigh\hlt{t}ing factors.
 
%
This approach also exploits the visual potential of SOMs which can be represented as 
bi-dimensional grids. In such grids, each cell 
 represents a group of nodes, and the size of the cells 
 is
proportional to the number of observations associated with it.
In this way the authors are able to represent the size of the communities, the distribution of topics and the links on the same 2-dimensional representation. 

\citet{Dang2012} propose an extension of the Louvain method with a modification of modularity to include the similarity of the attributes in the community discovery process. This is given by
\begin{equation}\label{eq:dang_distance}
Q=\sum_{C\in\mathbf{C}}\sum_{i,j\in C}\left(\alpha\cdot S\left(i,j\right)+\left(1-\alpha\right)\cdot simA\left(i,j\right)\right),
\end{equation}
where $\mathbf{C}$ indicates the set of graph partitions, 
$S\left(i,j\right)$ represents the strength between two nodes (computed as in the original definition of modularity), $simA\left( i,j\right)$ is a similarity function based on attributes $i$ and $j$ and
can be adapted according to how the attributes are represented. $0\leq\alpha\leq1$ is a weigh\hlt{t}ing factor.

In general, for parametric methods an important question is how to choose $\alpha$. According to the authors of these methods clusters are stable against small changes in the parameter.
\citet{Dang2012} also propose a way to estimate $\alpha$, and kernel-based approaches support automated parameter tuning \citep{villavialaneix2013}. Depending on application, analysts may also set $\alpha$ to emphasize attribute homophily or connectivity. However, more case studies and future independent analyses will be welcome.

\subsection{Walk-based approaches}

A \emph{random walk} on a possibly infinite network is a stochastic process where a walker goes from node to node by choosing a target neighbor at random at each step 
\citep{noh2004random}. 
In the clustering context  walk models are used to estimate vertex distances on attributed graphs. In accordance with this distance, $k$-means-like approaches attract \textit{close} nodes around the predefined $k$ centroids in order to aggregate the members of the communities.

\citet{Zhou2009} define a random walk process on graphs like the one in Figure~\ref{fig:attribute_augmented_graph}. The result is that the more
attribute values two vertices share, the more paths via the common attribute nodes exist. 
In this way random walks can be used to measure vertex proximity through both the structural links and the compositional links.

In the Connected $k$ Centers method proposed by \citet{Ge2008}
the walk strategy is a simple breadth-first search (BFS) defined for graphs like the one in Figure~\ref{fig:attributed_graph_2}, where the feature vector is also used to determine the next visited node. 
This method implements the $k$-means algorithm using walks to compute distances: first, it picks $k$ random nodes as cluster centers, second, all the nodes are assigned to one of the $k$ clusters by 
traversing the graph using BFS; third the centroids of the clusters are recalculated.
The second and third steps are repeated until there are no 
further changes in the clusters' centroids.

\subsection{\hlt{Methods based on statistical inference}}

\hlt{Statistical inference is the process of drawing properties of datasets from a set of observations in a model and then inferring predictions about a larger population represented by the sample. In this section, and according to the classification provided by} \citet{Fortunato2010}\hlt{, we focus on two types of methods: the ones using generative models, as an intermediary step or in a pure manner to mix attributes and links in a unified model, and the ones using stochastic block models.}

\hlt{Many studies focus on the task of clustering networks of documents. Here, every document can be seen as a node characterized by a complex attribute defined by the words contained in the document. For example, }\citet{Li08}\hlt{ propose a clustering method to find communities in a large-scale document corpus exploiting both the document content (the words), and their references/citations.  They use statistical inference as an intermediate step to find hidden topics to further manipulate the documents.
The general principle 
is to find community cores and then include their members. The detection of cores identifies the documents that are frequently co-referenced and may play the role of community seeds. 
A second phase merges the initial cores according to their topic similarity in order to improve the core consistency. The authors use here the well-known text-mining method called Latent Dirichlet Allocation (LDA) to find topics. LDA is a generative topic model so that unobserved or latent topics have probabilities to generate various observed words. A Bayesian inference finds the best fit of the model to the observations  through likelihood maximization.
Finally, the third step is to affiliate the remaining documents to the clusters. 
This affiliation propagation process may lead to misclassified documents and a final step removes false hits.}

\hlt{LDA is also used by }\citet{Liu2009} and \citet{Cohen2011}\hlt{ but as a central approach and in an extended manner to identify latent groups.}
\hlt{The Topic-Link LDA model defined by }\citet{Liu2009}\hlt{ is a generative model considering topics, membership of authors and 
link formation between pairs of documents exhibiting both topic similarity and community closeness. The inference is designed to regularize the topic information when inferring the hidden communities and vice versa. The authors maximize likelihood using an expectation-maximization algorithm and demonstrate their unified model on three different tasks: topic modeling, community detection and link prediction in blogs and CiteSeer datasets. For the community detection task, we would highlight here an interesting remark. Their approach offers a meaningful investigation of how content similarity and community similarity contribute to the formation of links. They are able to reveal that author membership has a much stronger effect on link formation between blog posts in political domains than technical papers. They also show that the topic dimension plays a more important role than the community similarity in blog citing.}
\citet{Cohen2011}\hlt{ also address the problem of link modeling and combine two popular methods: block modeling and LDA. }

\citet{Xu:2012:MAA:2213836.2213894}\hlt{ propose a community detection model that is transformed into a statistical inference problem. Authors start by defining a generative Bayesian model that produces a sample of all the possible combinations of a graph, defined by its adjacency matrix $\mathbf{X}$, a matrix of features $\mathbf{Y}$ and a vector $\mathbf{Z}$ containing the assignation of each node to one out of $k$ groups, i.e., a partition of the graph. This model produces a conjoint probability $p\left(\mathbf{X},\mathbf{Y},\mathbf{Z}\right)$. The idea is thus to find a partition $\mathbf{Z}^{*}$ such that $\mathbf{Z}^{*}=\arg_{\mathbf{Z}}\max p\left(\mathbf{Z}\mid\mathbf{X},\mathbf{Y}\right)$.}

\hlt{These techniques are very attractive to mix both attributes and topology into the same model, but unfortunately the optimization process to estimate the parameters of the likelihood is often costly. In addition, they do not rely on the definition of any distance, and the choice of the \emph{a priori} distributions in the statistical models requires a non-trivial expertise.}

\subsection{Subspace-based methods}

Some of the clustering approaches reviewed so far share the belief that a
carelessly usage of all the available 
attributes may lead to poor clusterings.
This is the case, e.g., in the work by \citet{villavialaneix2013}. 
We have already recalled the phenomenon called \textit{curse of dimensionality} in Section~\ref{weights-modification}: when the number of attributes is large the difference in the distance between two random pairs of data points (actors, in this case) tends to zero. This phenomenon motivates the development of clustering approaches focused on the identification of the discriminative attributes to produce well separated clusters. This general approach is known as \emph{subspace clustering}, and has been also applied to the case of node-attributed graphs.
Subspace clustering methods 
are designed to select the `best' subsets of dimensions. They search the projections of the data in different dimensions and identify clusters that are relevant \emph{locally} to some of these subspaces. 

Subspace clustering is interesting because it may reveal groups that would not be detected considering the entire set of attributes. However finding relevant projections is computationally hard. The final choice of which groups to keep is also costly and requires an optimization step combining the best size, density, entropy, dimensionality and any other relevant quality function (see Section \ref{sec:eval_intro}).
Moreover, as each cluster is relevant in its own subspace, 
this has the effect of producing overlapping clusters and requires additional efforts to control the redundancy ratio between them.

One semi-automated approach to identify relevant subsets of attributes has been presented by \citet{Cruz2011a}, where the authors 
propose a framework
helping human analysts to manually select their preferred compositional perspective. The choice of the subset of attributes is given explicitly as an input to an automatic clustering process.

Differently, \citet{gunnemann2013efficient} propose a completely automated method to efficiently combine subspace and subgraph clusters. 
In particular, they use their former GAMer method to extract an exhaustive list of candidate clusters, but apply a different final selection of the clusters to be returned to the user. The GAMer method greedily selects the clusters that locally optimize a quality measure. Here, they propose a solution based on global optimization, maximizing the sum of the clusters' qualities under redundancy constraints. 
The overall complexity of this definition of clustering is \#P-hard\footnote{This is the complexity of some hard counting problems, and implies that an exact solution to this problem cannot be currently computed in acceptable time}. Therefore, the authors propose a heuristic that, for example,
produces a clustering of the whole \hlt{DBLP} database\footnote{133 097 nodes; 631 384 edges; 2 695 attribute dimensions. Available at: http://dblp.uni-trier.de} in about 7 hours with commonly available hardware. 
They also show that the quality remains comparable to the greedy solution computed by GAMer in terms of F1 value and density. 

The time complexity of subspace clustering approaches is notoriously high, 
but the discovery of dense subgraphs in selected subspaces can be valuable. 
However, the high number of required input parameters (minimum cluster size, dimensionality, density, redundancy) can have a negative impact on the practical usability of these methods.
Finally, as we will see in \hlt{S}ection \ref{sec:eval_intro}, the evaluation of attributed graph clusters in general is still under study, and maybe more for overlapping ones where no ground truth exists.

\subsection{Other methods}

Other works directly extend well-known and efficient graph-based methods. 
\citet{Cruz2011b} extend the Louvain method \citep{Blondel2008} 
introducing a local minimization of the \textit{entropy} generated by the attributes between the modularity optimization and the community aggregation steps.
\citet{Dang2012} also extend the Louvain method in a similar way, by optimizing at each iteration the linear combination of the classical modularity and a new modularity based on the attributes.

\citet{conf/sdm/AkogluTMF12}\hlt{ propose a method to identify cohesive groups in attributed graphs composed of $n$ nodes each described by a feature vector. In this case, the attributes are binary, i.e., a node either has or not certain attributes.}
\hlt{The algorithm uses the adjacency matrix $\mathbf{A}_{n\times n}$ of the graph and a matrix $\mathbf{F}_{n\times f}$ representing the assignation of features for each vector. The main underlying idea is to find $k$ groups of nodes using the structural information and $l$ groups using the feature information.}
\hlt{The cost function is based on the encoding of the matrices $\mathbf{A}$ and $\mathbf{F}$ as well as the configuration of the clusters, 
where the encoding uses the approach proposed by }\citet{Risannen1983}.

\citet{Barbieri:2013:CCD:2433396.2433403}\hlt{ present an approach using the notion of information cascades, and in particular the idea that an information cascade is more likely to occur within a community rather than between communities. Thus, they use a given set of information cascades to build a probabilistic model named Community-Cascade Network (CCN). To learn the parameters of the model authors use an expectation-maximization approach, which however has been reported to be computational expensive.}

\citet{Ruan:2013:ECD:2488388.2488483}\hlt{ also propose a content- and structure-based community detection algorithm called CODICIL. The algorithm starts by creating an edge set with the structure and a graph generated from the similarity of the nodes, i.e.,  the final edge set will contain the original structure plus edges derived from obtaining the top $k$ most similar neighbors for each node. This similarity is calculated using the cosine distance between the TF-IDF vector from the content of each node. Then, this new graph is sampled to select certain relevant edges and, at last, this sampled graph is clustered using a classic graph clustering technique.}

\hlt{Finally, 
some approaches focus on the discovery of significant patterns, such as association rules or regular structures in graphs. Significant examples are the works by }\citet{Moser2009}, \citet{Silva2010}, \citet{atzmueller2011} and \citet{Pool:2014:DCD:2611448.2517088}\hlt{, focusing on mining descriptive community patterns and allowing the analysts to understand the structure of frequent subgraphs around topics which may be useful in scenarios like fraud detection or counter-terrorism. Differently from graph partitioning methods, frequent patterns can overlap and do not necessarily cover the entire dataset.}

\section{\hlt{Practical aspects}}
\subsection{Evaluation}\label{sec:eval_intro}

Comparing the quality of two clusterings is a fundamental capability. It can be used to choose among alternative algorithms, inside a single algorithm as a stopping condition or as a guide to choose the next step in a so-called \emph{greedy} \hlt{approach}, making an assignment that maximizes the quality improvement.
However, evaluating clustering algorithms is an open problem, even when graphs \hlt{without attributes} or even tabular data are involved. This has been clearly discussed in recent surveys by \citet{Schaeffer2007} and \citet{Fortunato2010} where the identified problems not only concern the ambiguous and personal definition of \emph{good cluster},
but also the need for results that are easier to interpret and use, benchmark datasets and quality functions to explain why a clustering is regarded as good or not.

\hlt{W}hile evaluating graph clustering is a hard and open problem even when no attributes are present, 
several measures to evaluate graph clusterings have been proposed, and some have been extensively applied. Therefore, without claiming that these measures represent the final or only solution to the problem, in this section we start from them as an existing way of evaluating graph clustering and focus on what we need to add when we deal with attributed graphs.

The main additional aspect to consider when attributed graphs are involved is the co-existence of multiple objective functions. Having a description of the data that includes both structural and compositional aspects, we may have sets of nodes that are very similar according to their attributes \hlt{but} disconnected \hlt{from} each other. Similarly, we may have well connected sets of nodes with \hlt{rather} heterogeneous compositional \hlt{attributes}. Both cases can be considered good clusters depending on the user requirements and while we would certainly prefer to identify sets of \hlt{nodes} making a good cluster with respect to all these aspects, we must accept the co-existence of multiple evaluation functions \hlt{---} or a \emph{multi-objective evaluation function}. 

In the \hlt{rest} of this section, we introduce relevant evaluation measures for different aspects involved in defining good attributed graph clusters. \hlt{In order to demonstrate their differences}, we apply these measures to a toy graph.

\subsubsection{Structural measures} \label{eval-struct}
Evaluating the quality of a clustering of a simple graph without node or edge attributes is a complex problem in itself. In this section, we will consider two different scenarios: evaluation with and without ground truth.

\paragraph{External evaluation measures.} When ground truth is available, the problem is reduced to computing similarity between two clusterings. Since we confront the found structures to externally provided class information, we call such similarity measures \emph{external evaluation measures}. 
These measures can be divided into two main groups: based on pair counting and based on information theory. We will briefly discuss the most typical representatives to give the readers an idea rather than a complete overview of the methods.

Given two partitions $\mathbf{C}_{u}=\{C_{u1},C_{u2},\ldots,C_{um}\}$ and $\mathbf{C}_{v}=\{C_{v1},C_{v2},\ldots,C_{vr}\}$ of \hlt{a} set of nodes, 
the pair-counting\hlt{-}based measures show the proportion of agreement between both partitions. 
\hlt{These measures have two requirements}: (1) the partitions are disjoint, i.e., $\bigcap_{C_{i}\in\mathbf{C}}C_{i}=\emptyset$, and (2) all elements have the same weight \hlt{in} the clustering process.

The \hlt{Rand index (RI)} is one of the first approaches for comparing two partitions \citep{Rand1971}. It can be considered as an alternative \hlt{to} accuracy because it expresses the number of pairs of nodes that were placed within the same group in both partitions divided by \hlt{the number of} all node pairs. 
This comparison leads to a similarity function $c\left(\mathbf{C}_{u},\mathbf{C}_{v}\right)$ between partitions that is expressed as
\begin{equation}\label{eq:rand-index}
c\left(\mathbf{C}_{u},\mathbf{C}_{v}\right)=\sum_{i<j}^{n}\frac{\gamma_{ij}}{\binom{n}{2}},
\end{equation}
where
$$
\gamma_{ij}=\begin{cases}
1 & \text{ if } \exists k,k' : x_{i},x_{j}\in{C}_{uk}\wedge x_{i},x_{j}\in{C}_{vk'}\\
1 & \text{ if } \exists k,k' : x_{i},x_{j}\notin{C}_{uk}\wedge x_{i},x_{j}\notin{C}_{vk'}\\
0 & \text{ otherwise. }
\end{cases}
$$

The agreements between partitions $\mathbf{C}_{u}$ and $\mathbf{C}_{v}$ can be summarized using a contingency matrix as presented in Figure \ref{fig:cont_mat_example}.
In this matrix, $n_{ij}$ is the number of agreements while $n_{i\cdot}$ is the number of elements of the $i$th group from the $\mathbf{C}_{u}$ partition and $n_{\cdot j}$ is the number of elements in the $j$th group in the $\mathbf{C}_{v}$ partition.

\begin{figure}[htb]
\begin{tabular}{cc|cccc|c}
 & \multicolumn{6}{c}{$\mathbf{C}_{v}$}\\ 
 & Class & $v_{1}$ & $v_{2}$ & $\ldots$ & $v_{r}$ & $\sum_{i\cdot}$\\ \cline{2-7}
 & $u_{1}$ & $n_{11}$ & $n_{12}$ & $\ldots$ & $n_{1r}$ & $n_{1\cdot}$ \\ 
 & $u_{2}$ & $n_{21}$ & $n_{22}$ & $\ldots$ & $n_{2r}$ & $n_{2\cdot}$ \\ 
$\mathbf{C}_{u}$
 & $\vdots$ & $\vdots$ & $\vdots$ & $\ddots$ & $\vdots$ & $\vdots$ \\ 
 & $u_{m}$ & $n_{m1}$ & $n_{m2}$ & $\ldots$ & $n_{mr}$ & $n_{m\cdot}$ \\
\cline{2-7}
 & $\sum_{\cdot j}$ & $n_{\cdot 1}$ & $n_{\cdot 2}$ & $\ldots$ & $n_{\cdot r}$ & $n$ \\ 
\end{tabular}
\caption{A contingency matrix representing the agreements $n_{ij}$ between two partitions}\label{fig:cont_mat_example}
\end{figure}

Using a contingency matrix similar to the one presented in Figure \ref{fig:cont_mat_example}, \hlt{E}quation \ref{eq:rand-index} can be re-expressed \hlt{as} 
\begin{equation}\label{eq:rand-index-rev}
c\left(\mathbf{C}_{u},\mathbf{C}_{v}\right)=\frac{\binom{n}{2}-\left[1/2\left(\sum_{i}\left(\sum_{j}n_{ij}\right)^{2}+
\sum_{j}\left(\sum_{i}n_{ij}\right)^{2}\right)-\sum\sum n^{2}_{ij}\right]}{\binom{n}{2}}.
\end{equation}

Note that $c\left(\mathbf{C}_{u},\mathbf{C}_{v}\right)\in[0,1]$, i.e., it is $0$ when the partitions are dissimilar and $1$ when the partitions are identical. Later \citet{Hubert1985} introduced the \hlt{adjusted Rand index (ARI)} which is a version of the Rand index corrected for chance. The ARI is given by
\begin{equation}\label{eq:ari_index}
ARI\left(\mathbf{C}_{u},\mathbf{C}_{v}\right)=
\frac{\sum_{i=1}^{r}\sum_{j=1}^{s}\binom{n_{ij}}{2}-\left[\sum_{i=1}^{r}\binom{n_{i\cdot}}{2}\sum_{j=1}^{s}\binom{n_{\cdot
j}}{2}\right]/\binom{n}{2}}
{\frac{1}{2}
\left[\sum_{i=1}^{r}\binom{n_{i\cdot}}{2}+
\sum_{j=1}^{s}\binom{n_{\cdot j}}{2}\right]-\left[\sum_{i=1}^{r}\binom{n_{
i\cdot}}{2}\sum_{j=1}^{s} \binom {n_{\cdot j}}{2}\right]/\binom{n}{2}},
\end{equation}
where $n_{i\cdot}$, $n_{\cdot j}$ and $n_{ij}$ are values taken from the contingency matrix in Figure \ref{fig:cont_mat_example}.


Another common measure is the Jaccard index which \hlt{is given} by the ratio of \hlt{the} node pairs that were clustered together in both partitions and the node pairs clustered together in at least one partition \citep{Jaccard1901}. 

The second group of external evaluation measures uses \hlt{mutual information} (MI) between partitions, i.e., the information both partitions share. These measures are based on entropy and joint entropy of the partitions. Using the same contingency matrix presented in Figure \ref{fig:cont_mat_example}, the MI index is given by
\begin{equation}\label{eq:mi_index}
MI\left(\mathbf{C}_{u},\mathbf{C}_{v}\right)=\sum_{i=1}^{m}\sum_{j=1}^{r}\frac{n_{ij}}{n}\log\frac{n_{ij}/n}{n_{i\cdot}n_{\cdot j}/n^{2}}.
\end{equation}
This measure can be normalized \hlt{by} the joint entropy of the partitions \hlt{ensuring} that the MI \hlt{lies} within the interval $[-1,1]$ or $[0,1]$. Variations of this measure with different normalizing factors or adjustments with correction for chance are presented in detail by \citet{Danon2005} and \citet{Vinh2010}.


\paragraph{Internal evaluation measures.} 
\hlt{Without} ground truth, determining the quality of a clustering is based on its intrinsic characteristics. We refer to such measures as \emph{internal evaluation measures}. 
According to \citet{ben2008measures}, ``a clustering quality measure is a function that maps pairs of the form $(dataset, clustering)$ to some ordered set (say, the set of non-negative real numbers), so that these values reflect how good or cogent that clustering is." Some general properties for good quality measures have been proposed, such as scale invariance, monotonicity and richness \citep{ben2008measures,Laarhoven2013},
but in practice the problem depends on the purpose of the analysis.

\hlt{To assess quality, } \hlc{\citet{Gaertler2005}} \hlt{uses two functions, $f\left(\mathbf{C}\right)$ and $g\left(\mathbf{C}\right)$, to measure, respectively, the density and the sparsity of the clustering. These functions are combined as follows}
\begin{equation}\label{eq:general_quality_framework}
index\left(\mathbf{C}\right)=\frac{f\left(\mathbf{C}\right)+g\left(\mathbf{C}\right)}{N\left(G\right)},
\end{equation}
\hlt{where $N\left(G\right)$ is a normalization function for the index defined as $\max\{f+g\}$ over all clusterings} \citep{Brandes2008}. \hlt{Using the general index defined in Equation} \ref{eq:general_quality_framework}, \hlt{three different quality indices can be derived: coverage, conductance, and performance.

\emph{Coverage} $\gamma\left(\mathbf{C}\right)$ is a measure of the ratio of the intra-cluster weights to the total amount of edge weights:}
\begin{equation}\label{eq:coverage}
\gamma\left(\mathbf{C}\right)=\frac{\omega\left(E\left(\mathbf{C}\right)\right)}{\omega\left(E\right)},
\end{equation}
\hlt{where $E\left(\mathbf{C}\right)$ is the set of intra-cluster edges and $\omega\left(\cdot\right)$ is the sum of the weights of a set of edges. According to the general definition in Equation} \ref{eq:general_quality_framework}, \hlt{$f=\omega\left(E\left(\mathbf{C}\right)\right)$ and $g=0$.}

\hlt{\emph{Conductance} $\varphi\left(G\right)$ is a measure based on the observation that if a cluster is well connected, then a large number of edges have to be removed in order to bisect it. Thus, conductance $\varphi\left(G\right)$ of a graph $G$ is the minimum conductance value over all
cuts of $G$} \citep{Brandes2008}\hlt{ --- that is, the lowest possible value of the total weight of all edges between the clusters of a partition $\mathbf{C}$. 
Along with the \emph{graph conductance}, two other measures exist:
intra-cluster conductance $\alpha\left(\mathbf{C}\right)$ and
inter-cluster conductance $\delta\left(\mathbf{C}\right)$. \emph{Intra-cluster conductance} is the minimum conductance value over all induced subgraphs $G\left(C_{i}\right)$ while the inter-cluster conductance is the maximum conductance over all induced cuts $\left(C_{i},\overline{C_{i}}\right)$. Thus, given a cut $\mathbf{C}=\left(C,\overline{C}\right)$, according to} \citet{Brandes2008}, \hlt{the conductances $\varphi\left(C\right)$ and $\varphi\left(G\right)$ can be defined as follows:}
\begin{align}
\varphi\left(C\right)&=\begin{cases}
	1, & C\in\{ \emptyset,V\} \\
	0, & C\notin\left\{ \emptyset,V\right\}\wedge\overline{\omega\left(\mathbf{C}\right)}=0\\
	\frac{\overline{\omega\left(\mathbf{C}\right)}}{\min\left(a\left(C\right),a\left(\overline{C}\right)\right)}, & \text{otherwise}
	\end{cases}\label{eq:conductance1} \\
\varphi\left(G\right)&=\min_{C\subseteq V}\varphi\left(C\right),\label{eq:conductance2}
\end{align}
\hlt{where $a\left(C\right)$ is the sum of the weight over all edges adjacent to $C$. It is expressed as}
$$
a\left(C\right)=2\sum_{e\in E\left(C\right)}\omega\left(e\right)+\sum_{f\in E\left(C,\overline{C}\right)}
\omega\left(f\right).
$$
\hlt{The intra-cluster conductance of a partition $\mathbf{C}$ is defined as}
\begin{equation}
\alpha\left(\mathbf{C}\right)=\min_{i\in\left\{ 1,\dots,k\right\}
}\varphi\left(G\left(C_{i}\right)\right),\label{eq:intraconductance}
\end{equation}
\hlt{while the inter-cluster conductance of a partition $\mathbf{C}$ as}
\begin{equation}
\delta\left(\mathbf{C}\right)=
	\begin{cases}
		1, & if\,\mathbf{C}=\left\{ V\right\} \\
		1-\max_{i\in\left\{ 1,...,k\right\} }\varphi\left(C_{i}\right), & \text{otherwise.}
	\end{cases}
	\label{eq:interconductance}
\end{equation}

\hlt{In order to express the preceding indices in the form of the general framework from Equation} \ref{eq:general_quality_framework}, \hlt{we set $g=0$ for intra-cluster conductance, $f=0$ for inter-cluster conductance and $N=f+g=1$ for both cases. }

\hlt{\emph{Performance} defines the quality of a partition based on the ``correctness'' of the classification of a node pair. 
The density function $f$ counts the number of edges within all clusters while the sparsity function $g$ counts the ``nonexistent edges" between clusters
}\citep{Gaertler2005},\hlt{ that is, the number of \emph{not} connected pairs of nodes among all clusters. The definitions are}
\begin{equation}
	\begin{array}{l}
		f\left(\mathbf{C}\right)={\displaystyle \sum_{i=1}^{k}\left|E\left(C_{i}\right)\right|}\\
		g\left(\mathbf{C}\right)={\displaystyle \sum_{u,v\in V}^{k}\left[\left(u,v\right)\notin E\right]I_{i,j}\left(u,v\right)},
	\end{array}
	\label{eq:performance1}
\end{equation}
\hlt{where the function $I$ is defined as:}
\begin{equation}
	I_{i,j}\left(u,v\right)=
	\begin{cases}
		1, & u\in C_{i}\wedge v\in C_{j},i\neq j\\
		0, & \text{otherwise}

	\end{cases}
	\label{eq:indicating}
\end{equation}
\hlt{Finally, performance as presented by} \citet{Brandes2008} is
\begin{equation}
	\textrm{perf}\left(\mathbf{C}\right)=\frac{f\left(\mathbf{C}
\right)+g\left(\mathbf{C}\right)}{\frac{1}{2}n\left(n-1\right)},
	\label{eq:performance2}
\end{equation}
\hlt{where $n$ is the number of nodes of the graph.}

\hlt{A comparison of clustering algorithms and measures has been provided by} \citet{Leskovec2010}\hlt{, and more details concerning the limitations of these measures can be found in the works by} \hlc{\citet{Gaertler2005} and \citet{Brandes2008}}.

Other candidates for a quality measure are \emph{density} and \emph{modularity} \citep{Newman2004,Fortunato2010}, and they can also be directly optimized instead of just being used as evaluation functions. We will not add additional details about modularity, that has already been described earlier in the article.

\begin{figure}[ht]
\centering
\includegraphics[width=.7\textwidth]{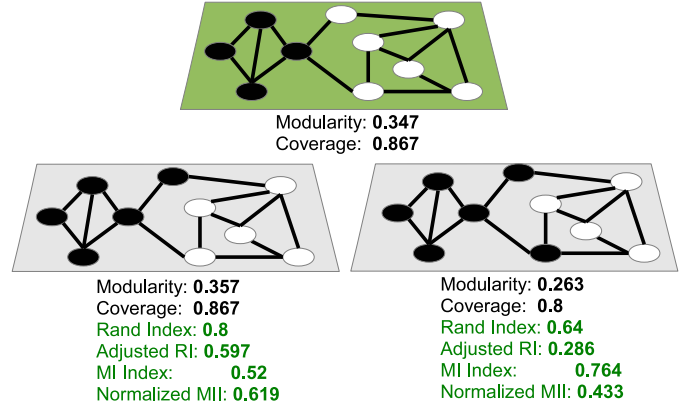}
\caption{Two graph partitions (lower graphs) and ground truth (upper graph): the values of some internal (modularity, coverage) and external evaluation measures (computed according to the ground truth) are shown}
\label{fig:eval-measures}
\end{figure}

In Figure \ref{fig:eval-measures}, some sample measures are illustrated on two alternative partitions of the same graph.

%

\subsubsection{Edge-attributed graph clustering} \label{eval-edge}


Only a few works have proposed evaluation measures for multiple graphs.
The measure introduced by \citet{Mucha2010a} takes into account both the pairs of nodes and the pairs of graphs --- this approach has already been described in Section \ref{modularity-based}. 

A different approach is given by \citet{Boden2012}. In the spirit of subspace clustering, a set of ``interesting" non-redundant clusters is sought. Candidate multidimensional clusters are considered to be all the node sets that are densely connected in every respective dimension (in all single layers that are contained in the cluster). From these, the result is selected by maximizing the quality sum $\sum_C{Q(C)}$ of all clusters while keeping the set of clusters non-redundant. Redundancy is computed as an overlap of edges of two clusters.
The quality function $Q(C)$ is meant to be specified by users since it is application-dependent. Nevertheless, the authors provide a default quality function which multiplies average density of the layers, size and dimensionality. Additionally, a minimum cluster size is set to 8 nodes and a minimum of 2 dimensions is required for each cluster. This evaluation measure is bound to a specific cluster model. \hlt{Moreover,} it is limited to finding multi-dimensional clusters that are clustered in all the single layers \hlt{at the same time (this results from the condition on the candidate clusters)}.

The problem of measuring distances between clusterings of graphs with weighted edges of multiple types is also tackled by \citet{Rocklin2011}.


\subsubsection{Node-attributed graph clustering} \label{eval-nodes} 
Node-attributed graph clustering approaches like the ones by \citet{Zhou2009}, \citet{Cruz2014tist} and \citet{Dang2012} use a combination of two measures: density $\delta$ for the structural part and entropy $\mathcal{H}$ for the attributes. Given a graph $G\left(V,E\right)$ and a partition $\mathbf{C}=\{C_{1},C_{2},\ldots,C_{k}\}$ of $G$, density is defined as:
\begin{equation}\label{eq:density}
\delta\left(\mathbf{C}\right)=\frac{1}{|E|}\sum_{C_{i}\in\mathbf{C}}|E\left(C_{i}\right)|,
\end{equation}
where $E\left(C_{i}\right)$ is the set of edges that start and finish in the $i$th community. \hlt{That is, density represents} the proportion of edges that lie \hlt{within the} communities \hlt{and a} higher density corresponds to a better clustering.

The term \emph{entropy}, used in several different contexts to \hlt{measure} the degree of disorder of a complex system, \hlt{indicates} the heterogeneity of the elements inside a cluster according to their attribute values\hlt{. It is given by}
\begin{equation}\label{eq:entropy}
\mathcal{H}\left(\mathbf{C}\right) = \frac{1}{|V|}\sum_{C_{i}\in\mathbf{C}}H\left(C_{i}\right),
\end{equation}
\hlt{where} $H\left(C_{i}\right)$ is the entropy of the $i$th \hlt{community} and is calculated as
$$
H\left(C_{i}\right)=-\sum_{j=1}^{r}p_{ij}\ln p_{ij}+\left(1-p_{ij}\right)\ln \left(1-p_{ij}\right),
$$
where $r$ is the number of \hlt{attributes} and $p_{ij}$ is the proportion of elements in the community $C_{i}$ with \hlt{the same value on} the attribute $j$. The objective \hlt{of the clustering} is to reduce the entropy which is equivalent to increasing the homogeneity of the partition.

Another validation technique is presented by \citet{Li08}. 
In this work, \hlt{documents are classified into ACM's 17 major computer science categories. This is a fuzzy classification that allows each document to belong to several categories. Thus, each document $d_{i}$ is assigned to a (17-dimensional) topic vector $z_{i}$ and then the documents are clustered into $K$ groups. Each group $C_{j}$ is further assigned to a topic vector $Z_{j}$.}

\hlt{The paper defines} a measure called $P\bar{C}S$ as
\begin{equation}
P\bar{C}S=\frac{PCS_{j}}{K},
\end{equation}
where $K$ is the number of communities, $PCS_{j}$ is
$$
PCS_{j}=\frac{\sum_{k:d_{jk}\in C_{j}}\eta\left(d_{jk}\right)}{n_{j}},
$$
\hlt{where} $n_{j}$ is the size of the community $j$ and 
$$
\eta\left(d_{jk}\right)=\begin{cases}
1 & \text{ if } z_{jk}= Z_{j} \\
0 & \text{ otherwise }.
\end{cases}
$$
Thus for each cluster $C_{j}$, the measure computes the proportion of elements $d_{k}\in C_{j}$ such that $z_{k}= Z_{j}$, i.e., how many documents within the community have a topic vector that is equal to the community's topic vector.

\hlt{In some cases it is possible to define the number and labels of the groups by hand as presented by }\hlc{\citet{Ge2008}}\hlt{ where authors compare the obtained partition with the expected one by counting the number of elements classified correctly by an algorithm. This approach is acceptable for small networks but becomes prohibitive for large networks with high dimensional feature spaces.}

When ground truth is available, it is possible to use validation methods such as Rand index or mutual information index. In this line, \citet{combe2012} define a framework for comparing the resulting partition with \hlt{the} ground truth. They use a contingency matrix (similar to the one presented in Figure \ref{fig:cont_mat_example}) created from the ground truth and a partition found by the tested algorithm. Then they calculate the proportion of nodes that were well grouped according to the ground truth.

\citet{Yang2009} use two validation approaches that are based on ground truth: the normalized mutual information (NMI), briefly described in Section\,\ref{eval-struct}, and the pairwise F measure (PWF). The PWF measure is given by \hlt{the} relation between \hlt{pairwise} \textit{precision} and \textit{recall}. This relation is
\begin{equation}\label{eq:pwf}
PWF=\frac{\left(1+\beta^{2}\right)precision\times recall}{\left(\beta\times precision\right)+ recall},
\end{equation}
where $\beta>0$ is a parameter used to favor either precision or recall. It is common to leave $\beta=1$. To calculate precision and recall, the following \hlt{expressions} are used
$$
\begin{array}{ccc}
precision&=&\frac{|S \cap T|}{|S|}\\
recall&=&\frac{|S \cap T|}{|T|},
\end{array}
$$
\hlt{where $S$ is the set of node pairs that are assigned to the same community and $T$ is the set of node pairs that have the same label.}



\subsubsection{A multi-objective evaluation approach}

In the previous sections we introduced several evaluation measures and we have seen that, in general, finding a good clustering of an attributed graph requires optimization of at least two objective functions. Therefore, there will always be a trade-off between compositional and structural dimensions. For node-attributed graphs, the objectives are the structural quality of the clusters (intra-cluster vs. inter-cluster edges) and the intra-cluster homogeneity of the node attributes. For edge-attributed graphs, the situation is more complicated since it is less obvious how to define a good clustering. According to \citet{Boden2012}, cluster candidates are well clustered in all of their dimensions, but 
this assumption could prevent the discovery of potentially useful clusters.

Another possible evaluation perspective consists in no longer checking if a clustering is good as a whole, but whether any specific interesting clusters are found.
In general, in order to evaluate a specific cluster in an attributed graph, one can take into consideration its structural quality, homogeneity of node attributes, size, dimensionality and novelty.
We can thus see these variables as different dimensions of a search space where each multidimensional point is a cluster. Good clusters can be selected based on custom settings of weights of the dimensions, or unweighted approaches like the Pareto front can be used to find all clusters that are potentially better than others according to any combination of these basic evaluation functions. 

For structural quality and node homogeneity, any measure from Sections \ref{eval-struct} and \ref{eval-nodes} may be selected. To assess novelty, we suggest to use one of the proposed measures of overlap, such as Jaccard index. The value of the maximum overlap can be returned as novelty. In this way, emerging clusters of minimal dimensionality are favored\hlt{, preventing} information overload.


\subsection{\hlt{Applicability}}


\hlt{Approaches preprocessing edge- or node-attributed graphs by reducing them to graphs without attributes normally keep the same asymptotic complexity of the clustering algorithm used after preprocessing. The exact complexity of the preprocessing phase depends on the data structure and the specific \emph{flattening} algorithm, but it is normally achievable in close-to-linear time on the size of the graph. As an example, edge-attributed flattening as described in Definition~}\hlc{\ref{def:flattening}} \hlt{and using a tree-based main memory indexing structure takes $\mathcal{O}(m\log{}m)$, where $m$ is the number of edges, that is, the average number of edges per edge type times the number of edge types. }

\hlt{As such, while not taking full advantage of the information represented by the different edge types, these methods can be applied to very large graphs using any of the existing efficient clustering algorithms reported e.g. by} \hlc{\citet{Coscia2011}}\hlt{, they are simple to implement and (with some variation in the flattening algorithm) can also be applied to directed and weighted graphs. However, in the case of weighted edge-attributed graphs, domain knowledge is necessary to decide how to merge weights on different edge-types. The conceptual problem of merging weights with different semantics as described by \citet{MagnaniSBP2013a}, e.g., the number of exchanged messages on an email layer and the duration of friendship on a social media network, emphasizes the deficiencies of single-layer approaches.}

\hlt{Similarly, for node-attributed graphs where the node attributes are flattened into edge weights before applying a community detection algorithm, the time complexity of the preprocessing step depends on the number of attributes and on the method used to compute how similar the nodes are. In case of \emph{matching similarity}, for each edge, the number of common attributes between the end nodes is computed which takes $\mathcal{O}\left(mf\right)$, where $f$ is the attribute space size and where in general $f\ll m$. In high dimensional spaces 
we expect that each node is described by a sparse vector and that allows for efficient methods such as growing self-organizing maps.
These methods, when coupled with efficient graph clustering, exhibit a near linear complexity. The other methods still take advantage of the sparse nature of the graph and thus, having less than quadratic complexity, are able to address large datasets.}

\hlt{With more integrated methods, such as subspace approaches or the one proposed by }\hlc{\citet{Ruan:2013:ECD:2488388.2488483}}\hlt{, the clustering process can reach a high complexity --- quadratic and more. But in general, for linear combination or walk-based methods, the complexity depends on the algorithm used for clustering the features, e.g., SOM or k-means among others, and whether the approach is global or local. The resulting process can still be practically used for reasonably large graphs, and graphs with hundred thousand nodes have been successfully processed in the reviewed works on subspace clustering. 

On the other hand, most of the community detection algorithms require the choice of parameters that control the output of the algorithm, for example the number of clusters $k$, the weight to emphasize the connectivity $\alpha$ or weighting variables for linear combination approaches, the number of iterations, statistical distributions for model-based methods, redundancy or heuristics in NP-hard subspace approaches; this sometimes requires major assumptions and domain knowledge about the data, which reduces their applicability. Only a few methods among the ones reported in this work are parameter-free, including the ones by } \hlc{\citet{Neville2003}, \citet{Cruz2011b} and \citet{conf/sdm/AkogluTMF12}}.

\hlt{Regarding the directionality of the edges, most of the methods described in this article rely on the application of existing approaches when the structural part of the graph must be analyzed, in which case any existing algorithm for directed graphs can be used. This evidently applies to the single-layer and weight-modification approaches, and is also the case for subspace methods, even if these last approaches may require some adaptation when specific algorithms have been hardcoded inside them. Methods based on extended modularity cannot be used without modifications on directed graphs, because they are based on the original definition of modularity which assumes undirected edges. However, they can be extended in the same way as it has been done by} \hlc{\citet{nicosia2009extending}} \hlt{for non-attributed graphs. With respect to node-attributed graphs, approaches based on linear combination can be straightforwardly used with directed edges as they are based on the computation of graph distances, that can be obtained on directed graphs as well. Similarly, walk-based approaches are naturally well suited to directed graphs.}

\hlt{As a final consideration, the works we have mentioned so far are all based on the general idea of clustering several dimensions at the same time; e.g., relationships, affiliation, competencies, socio-demographic features, among others. However, the information stored respectively in the attributes and in the edges may be uncorrelated and will not necessarily reinforce the same clusters. In practice, trying to merge several dimensions may result in failing to find any well separated clusters even when clusters exist under a single dimension.
An alternative approach is to run dedicated and specialized clustering steps for each dimension (structure, edge attributes, etc.), and then integrate the resulting partitions \textit{a posteriori} only if this leads to better clusters.
} \hlc{\citet{Cruz2013b} }\hlt{propose to manipulate the partitions with a contingency matrix where structural groups are in rows and compositional ones are in columns. The integration of the partitions relies on predefined strategies. Even if matrix manipulation may not seem user-friendly, this original proposal is interesting from another perspective: according to their objectives, the analysts can try different combinations without re-computing the basic partitions and thus potentially save computational costs.
}


\section{Open problems and discussion}

Attributed graph clustering is an active research area, and as such it presents a number of open problems.
In addition, being it an extension and combination of well established areas (graph clustering and multi-dimensional relational clustering), open problems can be classified into two main categories: \hlt{1) those} already present when single graphs are considered \hlt{(and the easier to identify) and 2)} those specifically related to the combination of structure and attributes. 



An example of the first category \hlt{pertains to} partitioning and overlapping algorithms. While the majority of graph clustering methods partition nodes into disjoint sets, many authors have pointed out that in real contexts individuals often belong to multiple communities. 
\hlt{Even without considering attributes, this has motivated the development of several methods,  such as the well-known clique percolation method by} \hlc{\citet{palla2005uncoveringOverlappingReference}}, or the ones by \hlc{\citet{nicosia2009extending}} and \hlc{\citet{Wang2011}} \hlt{where extended versions of modularity are used to evaluate overlapping clusters. In their recent paper,} \hlc{\citet{Xie2013}} \hlt{review the state-of-the-art in overlapping community detection algorithms, quality measures, and benchmarks for non-attributed graphs. They provide a framework to evaluate the performances of both the community-level and node-level detection, and conclude that this research field is still work in progress, as more than 70\% of the overlaps still remain uncovered.}
Other problems include how to measure the significance of overlapping nodes and how to interpret the 
resulting communities \citep{Xie2013}. Recently, \citet{Yang2013} have used node 
attributed graphs for detecting overlapping communities, stating that the 
resulting communities can be interpreted more easily by analyzing the 
attributes of the nodes belonging to each community. However, 
quality and interpretation issues are still open questions.

In general, like for other kinds of approaches, the presence of attributes introduces more parameters to be considered and requires the consideration of multiple aspects at the same time. However, in our opinion, when edge attributes are present, the dispute between partitioning and overlapping approaches should be reconsidered. In fact, overlapping is usually determined by participation in different networks: as an example, the same individual can be in her working team community, in her family community, in the community of her team mates at the fencing club, etc. This example suggests that if we can split our social network into a set of specialized networks (or, saying it in another way, if we can cluster our relationships into different classes), then we may find that \hlt{some} specialized networks only involve partitions. \hlt{However, this consideration should not be understood as a statement against overlapping methods}.

An example of the second category of open problems is the exponential explosion in the number of attribute value combinations to be considered during the clustering process. While this is a well-known problem in relational data mining, it is unknown in the domain of single graph clustering, and it is one of the main aspects reviewed in this article.
In Section \ref{sec:edges}, we hypothesized that clusters can emerge when a specific 
combination of graphs is considered, and disappear when more graphs 
are added to the model. In Section \ref{sec:nodes}, we discussed the notions of point 
of view and subspace clustering to counteract the fact that 
considering all the node attributes may lead to the \textit{curse of dimensionality} problem. 
Furthermore, beyond the \textit{quantitative} selection 
of a good subset of original data (which can be stated as a feature selection 
problem)%
, scientists will 
have to take into account \textit{qualitative} considerations: how to 
define the analysis context in order to decide how good a clustering is?
How to make this context understandable to an analyst without domain knowledge and usable by a domain expert without deep analytical skills? How to conceive efficient techniques to present multiple results in real-time?


The main problem related to the existence of multiple points of view which is peculiar of graphs with edge labels (and more in general multiple interconnected graphs) is the existence of a large number of views, where every view corresponds to a specific combination of values on the edges. Despite some promising attempts to address this problem, inspired by the field of sub-space clustering, in the authors' opinion this aspect deserves a lot more research to be able to apply clustering algorithms to real on-line social networks. Given the intrinsic computational complexity of the problem, a possible direction involves the consideration of domain knowledge to focus the cluster discovery process on promising combinations of dimensions.

Initial work in this direction by \citet{Cruz2013a} has defined control facilities to combine existing precomputed partitions. The objective is to offer tools to compare different approaches and visualize the results in a way that allows user feedback. The success of UCINET and --- more recently --- visual analytics software like Gephi and NodeXL is a sign that analysts are requesting such easy-to-apply tools. 
This requires advances focusing on usability, simplicity, efficiency and scalability, evaluation facilities such as comparison of methods, selection of relevant attributes and/or modeling. 
In fact, these research directions are as meaningful in an attributed-graph context as they are for non-attributed graphs.

Understandably, early works on attributed graph clustering have focused on finding static communities, which is a preliminary and necessary step to study their evolution. Here researchers can partially reuse the same approaches used to find evolving communities on simple graphs, in particular the comparison of nodes clustered at different timestamps to identify evolutionary steps like \emph{create}, \emph{merge} and \emph{split}. However, in the case of attributed graphs the evolution does not only regard the networks. 
The existence of multiple interconnected graphs and communities spanning some of them may also require a revision of the concept of evolutionary step.

A related problem that has generated a whole research sub-field in the realm of simple graphs is the study of network creation models. What are the forces leading to a specific network model exposing a modular structure? Rephrasing this question in the context of attributed graphs, \hlt{how can we} explain not only how some people have become densely interconnected, i.e., a cluster, but also why their attributes follow a specific value distribution and how these connections have developed in the different graph layers or edge types\hlt{?}

All the aspects mentioned so far highlight different levels of increasing complexity that we have to face when we consider attributes: the number of views to evaluate, the number of parameters to consider, e.g., in the evaluation functions, and the number of configurations of the system, e.g., the additional degrees of freedom in its evolution. A straightforward conclusion is that in the case of attributed graphs the applicability of forthcoming results may be strictly dependent on algorithmic advances, in particular regarding computational models like streaming, distributed, budget-based, approximate and incremental approaches enabling big data analysis.

\bibliographystyle{natbib}

\end{document}